\documentclass{bmcart}
\usepackage{makeidx} % allows index generation
\usepackage{graphicx} % standard LaTeX graphics tool when including figure files
\usepackage[caption=false]{subfig}
\usepackage{multicol} % used for the two-column index
\usepackage[bottom]{footmisc}% places footnotes at page bottom
\usepackage[cmex10]{amsmath}
\usepackage{placeins}
\usepackage[utf8]{inputenc}
\usepackage[english]{babel}
\usepackage{float}
\usepackage{rotating}
\usepackage{todonotes}
\usepackage{lineno}
\usepackage[pdftex]{hyperref} %Cannot use it as it clashes with EPJ .bst file
\showboxdepth=\maxdimen
\showboxbreadth=\maxdimen

\begin{document}

\begin{frontmatter}

\begin{fmbox}
\dochead{Research}

\title{Scholarly migration within Mexico: Analyzing internal migration among researchers using Scopus longitudinal bibliometric data}

\iffalse
\titlerunning{Scholarly migration within Mexico using longitudinal bibliometric data}

\author{Andrea Miranda-González\inst{1,2}%\orcidID{0000-0002-2845-2629}
,
Samin Aref\inst{2}%\orcidID{0000-0002-5870-9253}
,%\\
 Tom Theile\inst{2}%\orcidID{0000-0003-0573-9093}
 ,
 \and 
Emilio Zagheni\inst{2}%\orcidID{0000-0002-7660-8368}
}
%\authorrunning{A. Miranda-González et al.}

\institute{
Department of Demography, University of California Berkeley,
Berkeley, CA, USA
\email{andrea.mirgon@berkeley.edu}
\and
Max Planck Institute for Demographic Research, Konrad-Zuse-Str. 1, Rostock, Germany%\\
%\email{\{aref,theile,zagheni\}@demogr.mpg.de}
}
\fi

\author[
 addressref={aff1,aff2},  % id's of addresses, e.g. {aff1,aff2} 
 email={andrea.mirgon@berkeley.edu} % email address
]{\inits{AMG}\fnm{Andrea} \snm{Miranda-González}}
\author[
 addressref={aff2},
 corref={aff2},
 email={aref@demogr.mpg.de}
]{\inits{SA}\fnm{Samin} \snm{Aref}}
\author[
 addressref={aff2},
 email={theile@demogr.mpg.de}
]{\inits{TT}\fnm{Tom} \snm{Theile}}
\author[
 addressref={aff2},
 email={zagheni@demogr.mpg.de}
]{\inits{EZ}\fnm{Emilio} \snm{Zagheni}}

%%%%%%%%%%%%%%%%%%%%%%%%%%%%%%%%%%%%%%%%%%%%%%
%%   %%
%% Enter the authors' addresses here %%
%%   %%
%% Repeat \address commands as much as %%
%% required.  %%
%%   %%
%%%%%%%%%%%%%%%%%%%%%%%%%%%%%%%%%%%%%%%%%%%%%%

\address[id=aff1]{%  % unique id
 \orgname{Department of Demography, University of California Berkeley}, % university, etc
 %\street{},  %
 %\postcode{}  % post or zip code
 \city{Berkeley, CA},  % city
 \cny{USA}   % country
}
\address[id=aff2]{%
 \orgname{Laboratory of Digital and Computational Demography, Max Planck Institute for Demographic Research},
 %\street{Konrad-Zuse-Str. 1},
 %\postcode{18057}
 \city{Rostock},
 \cny{Germany}
}
\end{fmbox}

\begin{abstractbox}
\begin{abstract}
The migration of scholars is a major driver of innovation and of diffusion of knowledge. Although large-scale bibliometric data have been used to measure international migration of scholars, our understanding of internal migration among researchers is very limited. This is partly due to a lack of data aggregated at a suitable sub-national level. In this study, we analyze internal migration in Mexico based on over 1.1 million authorship records from the Scopus database. We trace the movements of scholars between Mexican states, and provide key demographic measures of internal migration for the 1996-2018 period. From a methodological perspective, we develop a new framework for enhancing data quality, inferring states from affiliations, and detecting moves from modal states for the purposes of studying internal migration among researchers. Substantively, we combine demographic and network science techniques to improve our understanding of internal migration patterns within country boundaries. The migration patterns between states in Mexico appear to be heterogeneous in size and direction across regions. However, while many scholars remain in their regions, there seems to be a preference for Mexico City and the surrounding states as migration destinations. We observed that over the past two decades, there has been a general decreasing trend in the crude migration intensity. However, the migration network has become more dense and more diverse, and has included greater exchanges between states along the Gulf and the Pacific Coast. Our analysis, which is mostly empirical in nature, lays the foundations for testing and developing theories that can rely on the analytical framework developed by migration scholars, and the richness of appropriately processed bibliometric data.
 \end{abstract}


% MSC classifications codes, if any
%\begin{keyword}[class=AMS]
%\kwd[Primary ]{}
%\kwd{}
%\kwd[; secondary ]{}
%\end{keyword}

\end{abstractbox}
%
%\end{fmbox}% uncomment this for twcolumn layout

\end{frontmatter}

\footnotetext{The reference to this article should be made as follows: {\scshape Miranda-González, A., Aref, S., Theile, T., and Zagheni, E.}
	\newblock Scholarly migration within Mexico: Analyzing internal migration among researchers using Scopus longitudinal bibliometric data.
	\newblock {\em EPJ Data Science}, (2020).
	\newblock doi: \href{https://doi.org/10.1140/epjds/s13688-020-00252-9}{10.1140/epjds/s13688-020-00252-9}.}

\section{Introduction}
\label{s:intro}

The academic exchange of ideas goes beyond physical borders. As such, many scholars are highly mobile, and their work contributes to technological and economic advances in a number of locations over the course of their academic lives. There is a growing body of literature that focuses on the migration and mobility of scientists and their impacts at the international level. However, even though the geographic distribution of scholars is both an outcome of regional disparities and a key source of inequality of opportunities for future generations, little is known about the drivers of these movements of researchers within country borders. Understanding these patterns can shed light on important regional deficits that identify areas of progress and opportunities for 
investment in human capital. From the public policy perspective, it is in the interests of regional governments to maintain a strong base of highly qualified scholars who can provide innovative and scientific solutions to public issues and collaborate with the private sector. In order to do so, regional governments should be aware of the underlying reasons for the migratory movements of researchers, and the associated sources of attraction at the national and the global levels. In order to identify these patterns, we propose an approach to study the internal migration of scholars using Scopus bibliometric data. We present our methods for measuring migratory movements, and discuss, as an illustrative case, the resulting network models of scholarly migration in Mexico.

There are a number of complementary theories of international and internal migration \cite{massey1993theories}, and there is a continuous effort to create a bridge between disciplines in order to build a more cohesive migration framework. Some theoretical migration studies have suggested that migration that occurs within a country depends on the stages of its development as a society \cite{Zelinsky1971}. Similar to the Demographic Transition theory \cite{coale1989demographic}, Zelinsky’s migration transition model identifies five phases in which spatial and time-specific characteristics (economic, social, and historical) determine mobility patterns \cite{Zelinsky1971}. Migration is studied by looking at different origins, destinations, and directions of migratory events. Migration estimates that provide a holistic view on the movements within a country would also facilitate the study of the temporal dynamics of a migration system from the perspective of a theoretical migration transition model.

The size and the impact of research in Mexico is perhaps not as well-known as those of other North American countries. According to SciVal 2010-2019 data \footnote{SciVal is a research profiling system and a web-based analytics solution provided by Elsevier \href{www.scival.com}{www.scival.com} (accessed on 25/06/2020)}, Mexico has over 200,000 researchers (comparable to Switzerland) and over 217,000 pieces of scholarly output (comparable to Singapore), with an average number of citations per publication \cite{colledge2014snowball} of 9.4 (comparable to China, Brazil, and Poland) and a field-weighted citation impact \cite{colledge2014snowball} of 0.91 (comparable to Japan and Iran). Despite these features, research in Mexico has been an under-studied case in the scientometrics literature. Historically, Mexico has deployed policies to attract foreign researchers through scholarships and professionalization, which placed it at the forefront of Latin American countries \cite{aupetit2006brain}. Existing work on Mexico has focused on research production and collaboration \cite{lancho2019science}; on particular fields of scholarship, such as computer science \cite{uddin2015scientometric}, physics \cite{reyes2016using}, and health sciences \cite{macias2013comparative}; and on Mexican researchers abroad \cite{marmolejo2015mobility}. 

Mexico is also a particularly relevant case for conducting exploratory analysis of internal migration, because most existing discussions on migration and Mexico are about migration from Mexico to the United States (US), even though between 2005 and 2010, inter-state and intra-state migration accounted for $3.5\%$ and $3.1\%$ of the population, respectively; whereas the rate of international migration was $1.1\%$ \cite{conapo_prontuario}. The migration of Mexican researchers abroad, particularly to the US, seems to be tied to differences in favorable conditions of the labor market \cite{lozano2015devaluacion}, and to changes in visa availability \cite{rodriguez2009migracion}. However, rather than focusing on a loss of talent, or \textit{brain drain} \cite{subbotin2020brain}, public policy can concentrate on harnessing the flows of researchers who come to (or return to) Mexico \cite{tuiran2013migracion}, and, similarly, on internal movers who can strengthen the domestic academic system. 

It remains unclear whether scholarly migration in Mexico has increased or slowed down in the last two decades as a result of special socioeconomic conditions, such as government spending on public institutions, social inequality, and the availability of alternative jobs in the private sector. Domestically, there have been ongoing efforts by the government to promote scientific research and development. For instance, a National System of Researchers (SNI in Spanish) has been created in order to track and reward academic and teaching contributions. Despite such policies, limited data availability has made evaluations of current internal mobility among researchers difficult and problematic.

This study intends to contribute to the literature in two main ways: first, by re-purposing bibliometric data to analyze internal rather than international migration; and, second, by exploring the migratory movements of scholars in Mexico. Although our substantive focus is on a specific country, the proposed methodological framework of re-purposing bibliometric data for internal migration and the proposed method of analysis are directly applicable to a broader context. We proceed, first, by describing the current state of tertiary education in Mexico and the available data sources in Section \ref{s:infer}. Then, in Section \ref{s:method}, we outline our methodology for adapting bibliometric data for migration research. We present our results on the demographic analysis in Section \ref{s:results}, and our results on network analysis in Section \ref{s:network}. Finally, we discuss and summarize our findings in Section \ref{s:discuss}.

\section{Inferring migration patterns for scholars from census data}
\label{s:infer}

\begin{figure}[ht]
 \centering
 \includegraphics[width=0.7\textwidth]{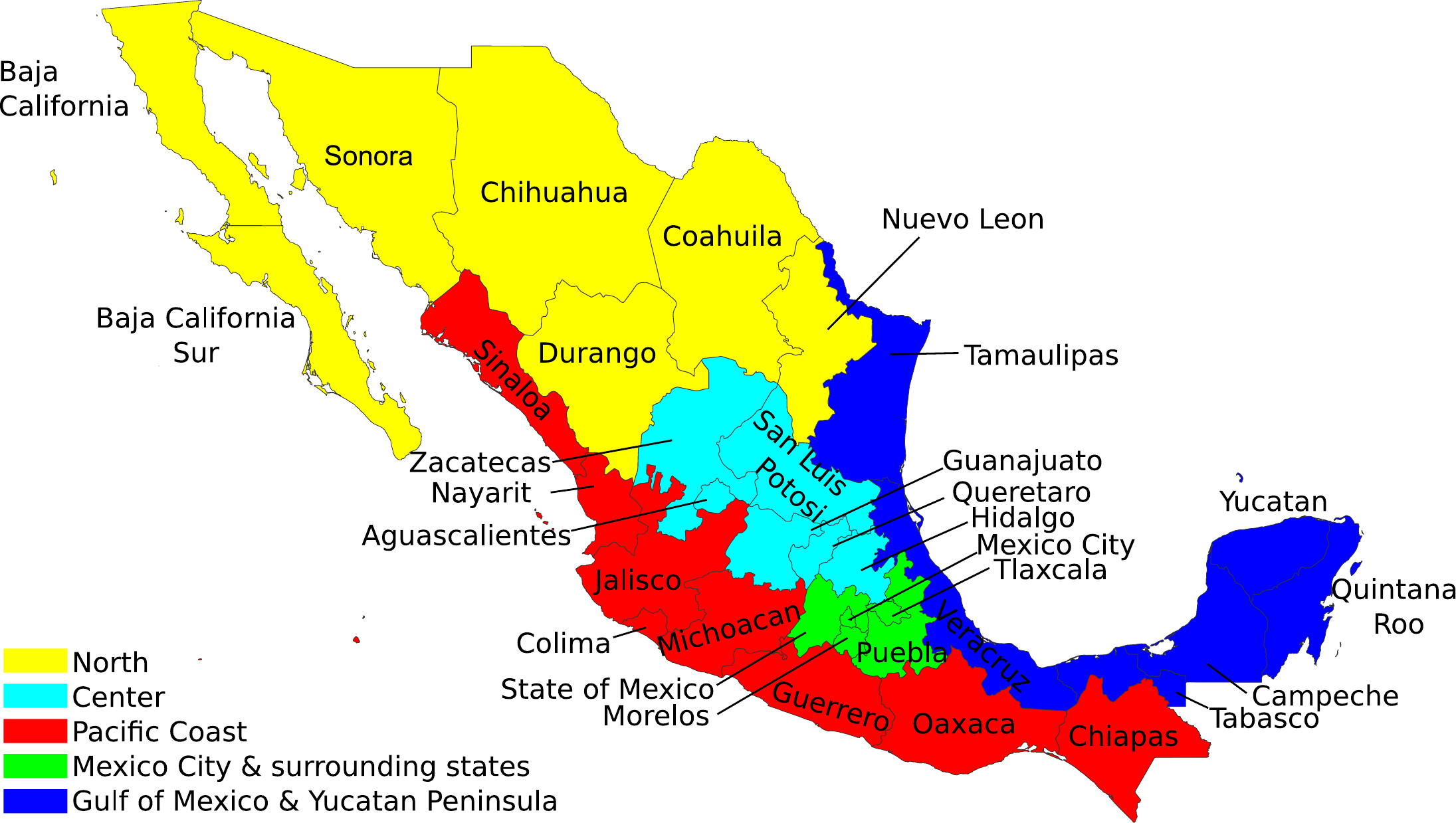}
 \caption{32 states of Mexico grouped into five regions}
 \label{fig:map}
\end{figure}

Mexico is composed of 32 states with specific natural endowments and economic infrastructures that enable them to accommodate over 126 million people. To better visualize the states where the movements occur, we include a map of Mexico in Figure \ref{fig:map}. In this study, we use two levels of aggregation: states and regions of Mexico. The latter are determined by the geographic, economic, and social similarities of the states. In total, we distinguish between five regions: i) Northern states along the Mexico-US border, ii) states in the Center with comparable economic status, iii) states along the Pacific Coast, iv) states surrounding Mexico City that share strong ties with the capital, and v) states benefiting from the industry and tourism of the Gulf of Mexico and the Yucat{\'a}n Peninsula.

The data sources that can be used to study specific groups of migrants are particularly limited. In Mexico, the National Institute of Statistics, Geography and Informatics (INEGI) collects information on the socio-demographic and geographic characteristics of the complete population every 10 years in the census, and through a large representative sample between censuses (five-year intercensus or population count). At an aggregate level, the National Council for Population (CONAPO) of Mexico provides yearly estimates and projections of internal net migration rates for the general population. However, the census data have some important limitations. As the census information on mobility yields transitions from the past state to the current state, we are unable to count the number of mobility events that could have occurred in between. Therefore, we cannot obtain the necessary numerators for computing internal migration rates, nor can we infer multiple movements over short periods for the same person.

\begin{figure}[ht]
 \centering
 \includegraphics[width=0.7\textwidth]{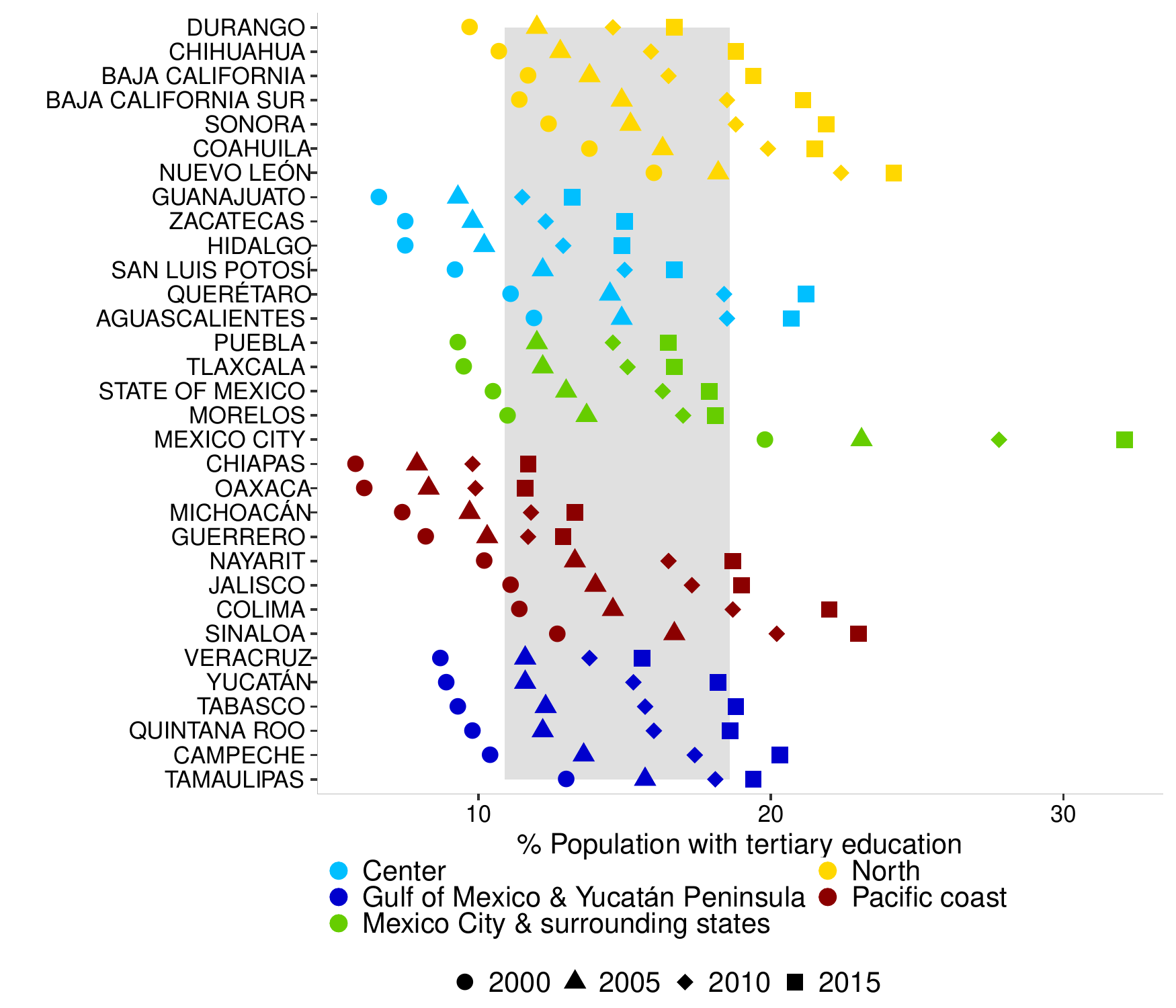}
 \caption{Share of population with higher education by states in Mexico. Gray area reflects the share of the total population with higher education between 2000 (lower-bound) and 2015 (upper-bound). Produced from INEGI aggregate level data (2000 census, 2005 population count, 2010 census, and the 2015 intercensus survey).}
 \label{fig:tertiary_pop}
\end{figure}

In the absence of bibliometric records, census data can be used to analyze the movement patterns of highly educated people, such as those with tertiary education. This is possible because census data contain information on an individual’s current residence relative to their last place of residence or their location from five years before, as well as their last attained educational level. On the educational front, Figure \ref{fig:tertiary_pop} shows the evolution of the share of population with tertiary education across different states in Mexico.\footnote{Aggregate data were obtained from INEGI \cite{inegi}, which defines higher education as technical or commercial studies with concluded high school studies, professional studies (bachelor’s degree or equivalent), specialties, master’s degree, or PhD.}

The national shares, in gray, range from $10.9\%$ in 2000 with an improvement up to $18.6\%$ by 2015. Overall, there has been an increase in the share of tertiary-educated people in every state. While all states have experienced growth in the share of the population with tertiary education, many still lag far behind the national average. This is the case for Chiapas, Oaxaca, Michoac{\'a}n, and Guerrero, all of which are states along the Pacific coast of Mexico. In several other states, such as in states in the North (in yellow), Quer{\'e}taro, Aguascalientes, Colima, and Sinaloa, the share of tertiary-educated people has increased considerably. Unlike in all other states, in Mexico City – which, in addition to being the capital, is an important center for social and economic activities, and is the home of many government offices – the share of the population with post-high school education has surpassed $30\%$, as of 2015.

\begin{figure}[ht]
 \centering
 \includegraphics[width=0.98\textwidth]{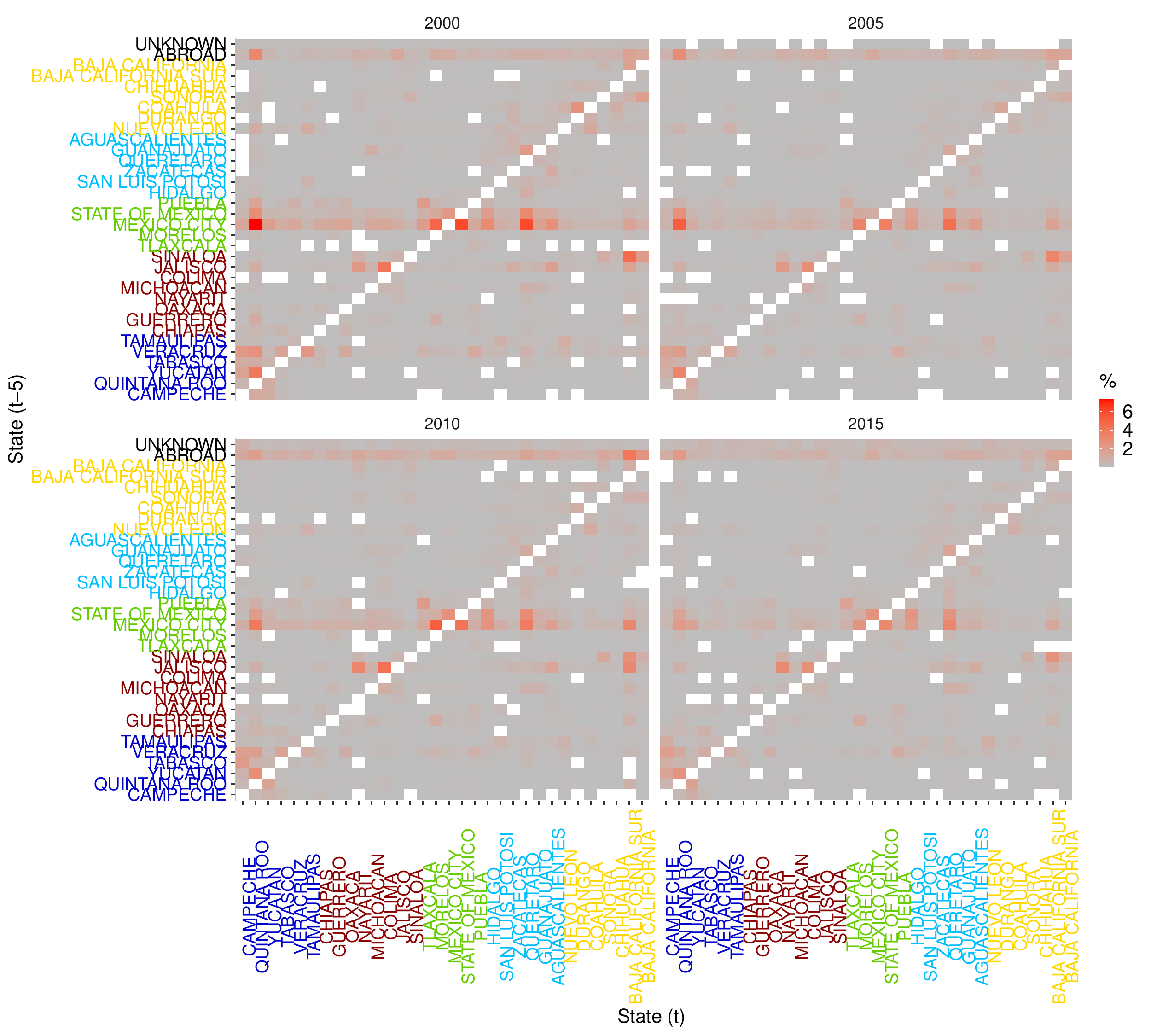}
 \caption{Share of population with tertiary education who were in a different state 5 years prior to 2000, 2005, 2010, and 2015. Colors of states refer to regions from Figure \ref{fig:map}. Produced from extracts of IPUMS International data (2000 census, 2005 Population Count, 2010 census, and the 2015 Intercensus Survey).}
 \label{fig:ipums}
\end{figure}

We use the census microdata to focus on the movements among the population with tertiary education. Using IPUMS International extracts from 2000 to 2015 on Mexican census data \cite{ipums}, we exploit the information on the location of residence five years prior among people who have a higher level of education. Figure \ref{fig:ipums} shows the fraction of individuals with tertiary education who moved between two states (those who left the y-axis state five years prior to the census year to move to the x-axis state) over the tertiary-educated population of the destination state.

We can see that while there are movements from most of the states, Mexico City appears to play an especially important role, as it was a common origin of tertiary-educated movers. Moreover, the movements appear to be regional, in particular within the states of i) the Gulf of Mexico and the Yucat{\'a}n Peninsula (in blue), and ii) Mexico City and its surrounding states (in green), which can be seen by looking at the concentrated red colors along the diagonal. However, using the census data to infer the migration rates of specific subgroups of the population proves to be difficult and severely limited by the five-year time span, even if the data are linked. On the other hand, the census provides information on important individual demographic characteristics, such as age and sex, which are not readily available from other sources like bibliometric data.

Based on these observations, we can state that using census data to analyze the movements of researchers would provide a general image of reduced movements of people with higher education that is not necessarily indicative of the movements of scholars. In other words, census data enabled us to observe the movements of tertiary-educated people, but not the movements of the research-active scholars who presumably contribute to the creation and dissemination of focused knowledge.

\section{Methodology of re-purposing bibliometric data for internal migration}
\label{s:method}

In order to analyze the international migration of researchers, many studies have relied on bibliometric databases. Recent studies offered proxies for place of residence \cite{czaika_high-skilled_2018}; provided information on bilateral international migration flows \cite{czaika_globalisation_2018}; offered a methodological framework for dealing with multiple affiliations \cite{robinson-garcia_many_2019}, analyzed the movements of highly mobile researchers and return migration flows \cite{aref2019demography}, and studied the impact of migration on different fields of scholarship in a country \cite{subbotin2020brain}. In particular, Scopus has been widely used to analyze international mobility \cite{moed2013studying,moed2014bibliometric,subbotin2020brain} due to its advantages compared to other bibliometric databases. For instance, Scopus provides a wider breadth of records for a number of disciplines \cite{falagas2008comparison} and offers a more reliable author ID, \cite{kawashima2015accuracy} which is considered to be more suitable for tracking the movements of individual researchers \cite{Aman2018}.

The availability of large-scale bibliometric data allows us to track the migration of researchers in a way that has not been possible with traditional sources of migration data, like censuses and surveys. Additionally, bibliometric data provide standardized data, which are suitable for comparative studies. The unit of the data is an \textit{authorship record}, which is the linkage between an author and a publication. The data that we use in this paper involve 1.1 million authorship records of scholars who have published with Mexican affiliation addresses in sources covered by Scopus. Using these data, we analyze information for over 252,000 researchers across 32 states of Mexico and the changes in their affiliation addresses over the 1996-2018 period. In the next subsections, we explain the pre-processing steps, which are the key parts of repurposing bibliometric data for an analysis of internal migration among researchers.

\subsection{Improving Scopus author IDs}

We use the Scopus author ID \cite{kawashima2015accuracy,Aman2018} to group the authorship records of individual researchers and to detect mobility events. While most author IDs correctly disambiguate individual researchers, and the Scopus author IDs are considered to be reliable for inferring mobility in the literature \cite{Aman2018}, the Scopus author identification system is not perfect, and there could be multiple distinct individuals sharing the same name (or similar names) who have been incorrectly allocated the same author ID. Author name disambiguation remains a challenge in scientometrics, although there have been many recent efforts to tackle it algorithmically \cite{tekles2019author,dangelo2020disambiguation}.

The first step of our pre-processing of the data involves treating the authorship records that are most likely to be impacted by the lack of accuracy in the Scopus author identification system. We extract all of the authorship records that are associated with more than 276 \footnote{While this number is not chosen through a rigorous scientific process, we remind the reader that the aim here is to consider a threshold for particularly suspicious authorship records whose size could vary depending on the quality of the author disambiguation method used by the provider of the bibliometric data. Given the very small fraction of such records in the Scopus data, this process would not change our substantive results and discussion. Nonetheless, we apply a healthy degree of skepticism to ensure a sound data analysis that starts with a treatment of outliers.} publications (an average of more than one publication per month during the 23 years of our analysis period). We implemented an approach inspired by the state-of-the-art methods in the literature \cite{dangelo2020disambiguation} to further disambiguate the extracted set of authorship records through a conservative approach of allocating revised author IDs. We explain our method briefly, while more information about the task of author name disambiguation can be found in \cite{dangelo2020disambiguation}.

Our algorithm takes the list of authorship records of a single suspicious author ID and compares every pair of authorship records on the list. For each pair, we check the similarity of the author name (by comparing characters), the number of shared co-authors (on the two authorship records), the article classification (discipline subjects of the publication venue), the funding organization or grant numbers, and the affiliation addresses. For each topic of pairwise comparison, we assign a score that is higher if the two authorship records have similar traits, and a score that is lower or negative if they are dissimilar. We sum the score into a single value, and calculate a distance matrix for all combinations of records in the list. We obtain clusters of authorship records using the \textit{agglomerative clustering} algorithm in the \textit{scikit-learn} software package \cite{scikit-learn}, and assign a revised author ID to each cluster of authorship records.

Among more than 250,000 distinct author IDs in our data, only 27 author IDs were associated with more than 276 publications. However, these 27 author IDs accounted for 15,229 authorship records associated with 4,207 distinct publications (note that some publications are shared between these authors). After applying our disambiguation algorithm, we identified 130 disjoint clusters of reasonably similar authorship records to which we assigned 130 revised author IDs, instead of the 27 IDs allocated by Scopus. There were two groups for which the Scopus author IDs were not changed. Upon manually investigating such cases, we found particularly prolific researchers who happened to have authored over 276 publications each. The majority of the 27 groups of authorship records were broken into smaller groups as multiple revised Author IDs were allocated to their records. We manually checked some of the cases, and found that they were indeed authorship records of multiple distinct individuals sharing the same name whose records were correctly disambiguated by our algorithm.

Our algorithm performs well for particularly suspicious outlier records selected from the end of the publication count distribution. However, running it on all of the data would not be helpful, because it was not designed to be a replacement for the Scopus author identification system. The conservative idea behind the algorithm is that by default, each pair of authorship records is from a distinct individual unless there is sufficient evidence (of similarity) to the contrary. Running such an algorithm on other groups of authorship records sharing the same author ID may incorrectly break the authorship records of a single researcher into smaller groups, especially if the person has visible points of change in their career; if, for instance, they have changed their academic name, discipline, and co-authors.

\subsection{Inferring states from affiliations using a simple rule-based algorithm}

A second and highly important step of our data pre-processing is aimed at extracting the state from the affiliation addresses of each authorship record. Affiliations are not standard, and they vary substantially. Scopus breaks down these affiliations into the following variables: name of an institutional unit or department, name of an institution, street address, city, postcode, and country. However, many of the fields may be missing. For instance, some authorship records may contain only the name of an institutional unit (e.g., a research group) without any address or postcode. Therefore, we first need to harmonize the text data by removing accents and special characters, and by standardizing common abbreviations. Then, we use a simple rule-based algorithm that sequentially looks for each of the affiliation variables in dictionaries where the cities, postcodes, and names of institutions are mapped to their respective states. Specifically, we use as dictionaries data from the Mexican Post Office \cite{postalmexico}; the ANUIES inventory of academic institutions \cite{anuies}; and CADIIP \cite{cadiip_2012}, an inventory of public institutions in Mexico. For addresses, the algorithm looks for the name or abbreviation of a state within the string. As a result, for each authorship record, we obtain up to five potential states based on the previous matching process. Finally, we define the state to be the mode of the five possibilities, as long as it is detected at least twice. In doing so, we exploit all the available geographic information per record, rather than relying on a single variable. For $75\%$ of the data, the states can be extracted reliably using the mode state outputs of this simple, rule-based algorithm. However, for the remaining $25\%$ of the authorship records, reliable states cannot be extracted using the rule-based method, and a more sophisticated method is therefore required to identify the states. In Subsection \ref{ss:develop}, we explain how we developed a neural network approach to address this issue; while in Subsection \ref{ss:using}, we discuss how this approach is applied to extract the remaining states.

\subsection{Developing a neural network and measuring its performance} \label{ss:develop}

We use the states obtained through the rule-based method as training data for developing a neural network for inferring the states of the remaining part of the data. We use the \textit{Keras} library \cite{keras} with a \textit{tensorflow} backend \cite{tensorflow} in the Python programming language.

\begin{figure}
 \centering
 \includegraphics[width=0.9\textwidth]{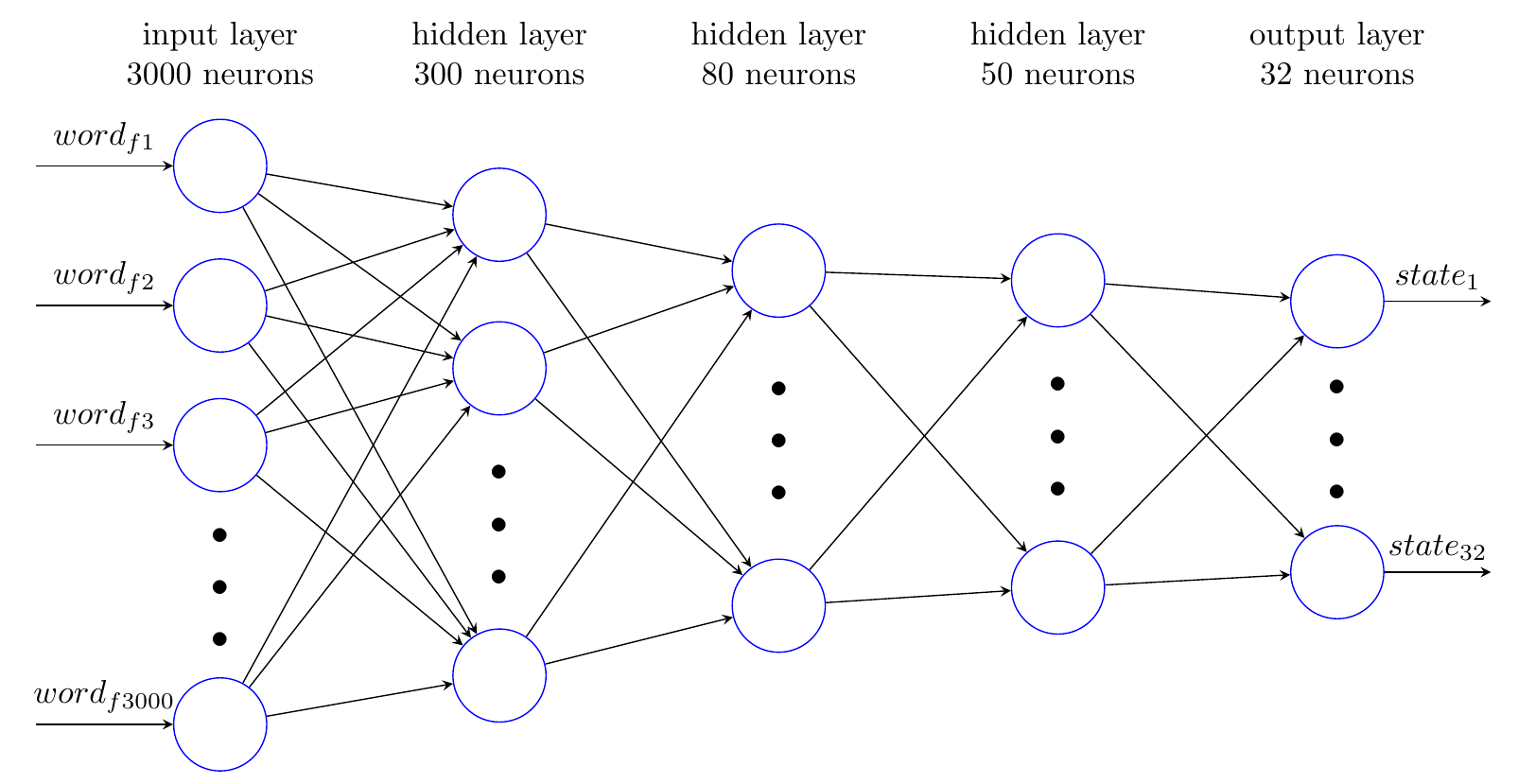}
 \caption{A schematic representation of the neural network used for inferring states from affiliations}
 \label{fig:neural}
\end{figure}

To assign a state to each row of the input consisting of a city, an institution, and an address, we adopted an approach commonly used in sentiment analysis literature \cite{mantyla2018evolution} for predicting whether a sentence has a positive or a negative sentiment. Initially, we merge the city, the institution, and the address into one string for each row of the input. Then, we convert this string into a \textit{feature vector} using a bag-of-words method with \textit{term frequency inverse document frequency} \cite{tfidf} as the normalized frequency of a given word in a given row of the input relative to the frequency of that word in the whole data set. We use the words with the 3,000 highest relative frequencies as the input layer of the neural network. The number of layers and the number of neurons of the neural network are depicted in Figure \ref{fig:neural}.

The layers are densely connected, and have a dropout rate of $0.25$. The neurons of the first layers use a rectified linear unit \cite{relu} as the activation function, while the output layer uses a \textit{softmax} activation function \cite{bishop2006pattern}. The softmax activation function converts the activation of the output neurons into relative probabilities. The state that is assigned to the output neuron with the highest activation gets selected as the predicted state. We fine-tuned the number of layers, the number of neurons, and the dropout rate by trial and error to achieve a high degree of accuracy on the test data, while keeping the network as small as possible to avoid over-fitting.

We use the records for which the states were already obtained by the rule-based approach as training data for the neural network. Additionally, we manually labeled the states of 2,448 authorship records that were not only unrecognizable by the rule-based approach, but were especially difficult to predict, as they failed on multiple rules. We added 1,574 ($70\%$) of those manually labeled records to the training set. Using a five-fold cross-validation, we obtained a median accuracy of $99.5\%$ on this dataset. We then tested the accuracy of the method on the remaining 674 ($30\%$) of the difficult-to-predict records, and obtained an accuracy rate of $96.2\%$. This could be considered a conservative evaluation for the accuracy of the neural network.

\subsection{Inferring states from affiliations using the neural network} \label{ss:using}

We used the neural network approach described in the previous subsection to predict the states of the records for which states could not be reliably obtained using our rule-based approach. We took additional steps to increase the overall reliability of the predictions. More specifically, we omitted a small fraction (less than $1\%$) of the total number of authorship records for which predicting a reliable state was particularly difficult even after using the neural network. Only the predictions with the lowest confidence score\footnote{The confidence score is the activation of the neuron with the highest activation in the softmax output layer.} were omitted from the dataset.

In summary, the states for $75\%$ of the authorship records were reliably extracted using a simple rule-based method. The states for a remaining $24\%$ of the authorship records were obtained using a neural network with a high degree of accuracy. Less than $1\%$ of the authorship records were discarded because their states could not be extracted in a reliable way using either of our two methods.

\subsection{Detecting moves based on changes in mode states across different years}\label{ss:modestate}

After extracting states for the authorship records, we obtained the most frequent state for each researcher in each year. Note that, in some cases, a researcher may have multiple modal states in a year. We consider a move only if the changes in affiliations are such that the modal state of a researcher changes and the previous modal state disappears. For example, let us assume that the modal state for a researcher in years 2001, 2003, and 2006 are [Morelos], [Morelos, Hidalgo], and [Hidalgo], respectively. Our move detection algorithm iterates over the years in which the author has publications (and therefore modal states). When the algorithm reaches 2003, the modal state changes – it is a double mode – but the previous state has not disappeared; therefore, we do not consider that modal state change a move. Instead, we assume that the researcher is still affiliated with Morelos. When we reach 2006, the mode has changed, and the previous mode has disappeared. Thus, a move from Morelos to Hidalgo is recorded by the algorithm. To infer the time of the move, we consider the mid-point of the two years being processed by the algorithm when the move is detected. In the example above, the move-year is recorded as the average of 2003 and 2006. When the move-year is not an integer, we round it down to the closest integer.

\section{Results}
\label{s:results}
\subsection{General attributes of scholars in Mexico}

One of the benefits of bibliometric data is that they enable us to conduct analyses at both the individual and the aggregate level. For example, we can obtain a profile of the median scholar in Mexico by looking at each scholar’s mobility status, and complement this information with migration rates for each state. The data resulting from the pre-processing discussed in Section \ref{s:method} allowed us to identify about 252,000 unique scholars in Mexico who were active during the 1996-2018 period. However, a large share (about $57\%$, or 146,000 authors) of these unique scholars are individuals who only have one authorship record, which prevents us from inferring their internal migration patterns. These one-time observations could be of scholars who moved abroad after publishing a single paper with a Mexican address, or of individuals who did not remain active in academia in Mexico. After removing these observations, about 96,000 authorship records are left, which cover 22,000 scholars who have moved at least once. That is, when we group individuals into moving and non-moving categories, the data show that only $22.8\%$ of scholars who have published more than once also moved between states during the 1996-2018 period.

Three available characteristics of scholars are their academic age, number of publications, and total citation count. The publication dates allow us to compute the number of years each researcher has been active. \textit{Academic age} is obtained by subtracting the year of an author’s first publication from 2019, in the spirit of \cite{aref2019demography}. In doing so, we assume that scholars have not finished their academic trajectory, and that the first year of publication can be thought of as the birth year of a scholar. Overall, the median academic age of the scholars in our data is eight years, but the mobile scholars have a median age of 13 years, while their non-mobile counterparts have a median age of eight years. Indeed, the mobile scholars appear to be active for longer, which could be an artefact of having to build a more solid track record in order to be eligible for opportunities to transfer from one institute to another. In terms of a crude productivity measure, the median mobile scholar has five publications, whereas the median non-mobile scholar who has published more than once has three publications. Additionally, we can obtain a total citation count by author for the available publications, and normalize it by the academic age. In doing so, we control for the longer academic careers of some scholars. The median mobile scholar received 2.25 citations per year, while the median stationary scholar was cited 1.5 times per year.

A potential explanation for both of these productivity measures is that the typical mobile scholar has a longer career because it has taken them longer to accumulate networks to migrate to another state. At the same time, there could be a bias towards detecting mobility from researchers with more publications. While the Scopus database contains authorship records up to a certain date, the profiles of researchers are continuously evolving. Thus, the measures can only be used to infer the presence of differences between these two groups, rather than the magnitude of such differences.

\begin{figure}[ht]
 \centering
 \includegraphics[width=0.75\textwidth, scale=0.5]{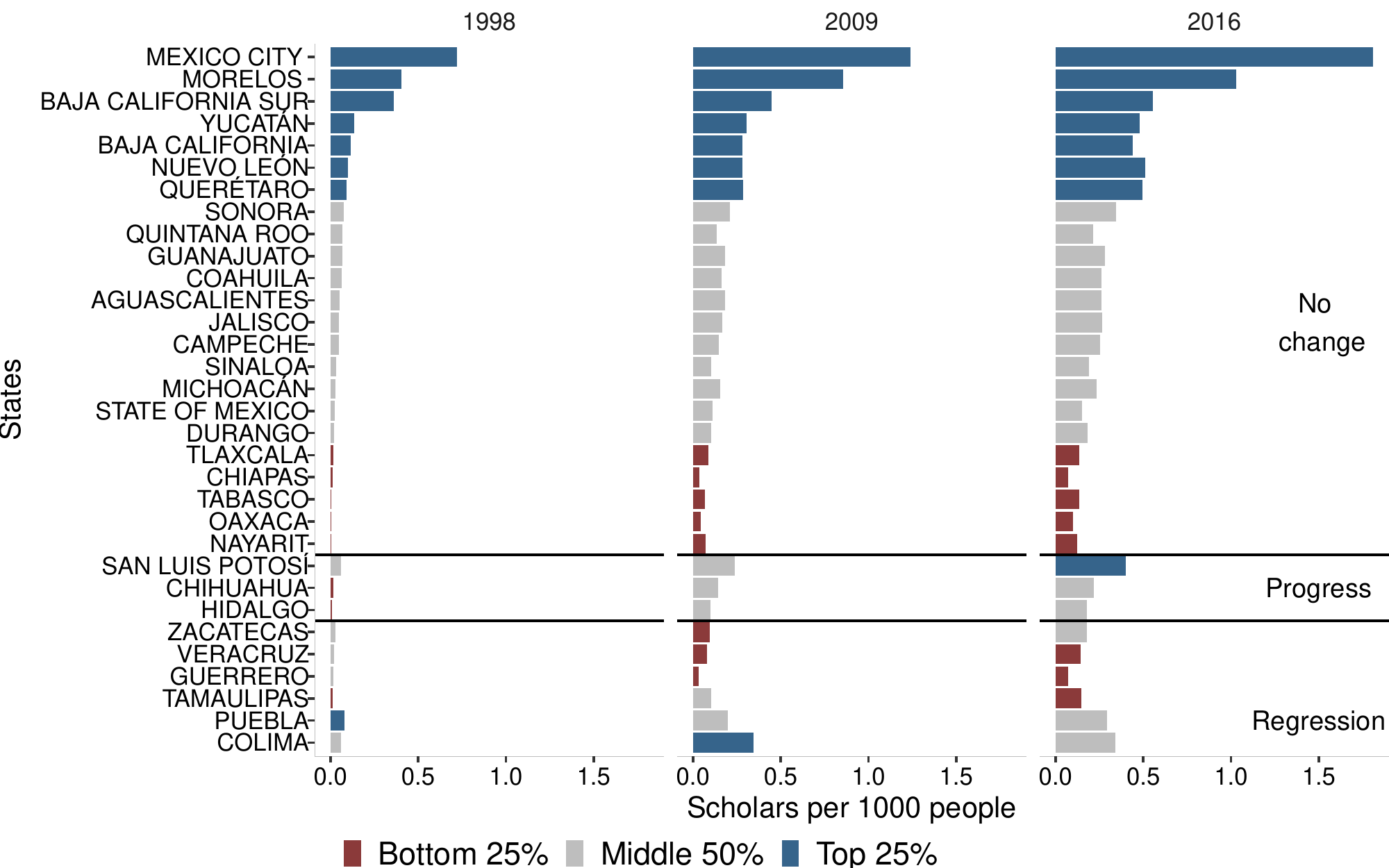} 
 \caption{Scholars per capita in Mexico. \newline The horizontal lines separate states that have experienced ``No Change", ``Progress" and a ``Regression" in their relative position across time. For instance, a state in the ``Progress" category may have a density in the bottom 25\% in 1998 but a higher density in the middle 50\% or higher later on. A state in the ``Regression'' category experienced a relative fall in its density of scholars: for example, by being in the middle 50 \% in 2009 and falling into the bottom 25\% in 2016. }
 \label{fig:scholar_pc}
\end{figure}

Scopus data allow us to obtain complete scholar counts per year once we aggregate the data per state. However, the publications may not occur in consecutive years, and the bibliometric data can only offer screenshots at specific times. In the absence of an observation, we must assume that each scholar remained in the same state over the last and following two years of a publication\footnote{Essentially, we fill backwards and forwards the two neighboring missing values around a publication year. Due to this padding, the effective sample when using scholar counts is reduced to 1998-2016.}. After all, obtaining positions in other states and moving is costly and difficult in academia. After these initial analyses, we aggregate counts of scholars, and use separate CONAPO midyear population estimates \cite{conapo} from each state to obtain the density of scholars, as seen in Figure \ref{fig:scholar_pc}.

For every year, each bar represents the number of scholars per 1,000 people in a given state, and its color indicates the state’s relative position in a given year, which allows for comparisons in the presence of population growth trends. Over time, there is an increase in the density of scholars per state that reaches a maximum of 1.8 per 1,000 residents of Mexico City. While many states remain in the same percentile, some states – San Luis Potos{\'i}, Chihuahua, Hidalgo – experience an increase in the number of scholars, and fall in the ``Progress'' band. There are still five states that do not fare as well as the rest of the states, even if their scholar density increases, and thus fall into the ``Regression'' category. Within this group, only Colima and Tamaulipas have shares of the population with tertiary education that are higher than the national value in the last available year (Figure \ref{fig:tertiary_pop}). When we look at the states with the highest densities (in blue) in 2016, we see Yucat{\'a}n which, coincidentally, has a smaller share of the tertiary-educated population than the national average in Figure \ref{fig:tertiary_pop}. Therefore, the finding that a state has many scholars does not necessarily imply that the state has high levels of education. Moreover, the differences in the factors that attract scholars seem to persist. For example, in 2016, the density of scholars in Chiapas continued to be lower than that of Mexico City.

\subsection{Measures of internal migration}
Our data set provides the history of movements over years, from which we can determine whether a scholar has relocated to a different state. As the time that it takes to conduct a research project and to publish the results varies by discipline, for many scholars, there is a gap of years between each publication – and, therefore, between each potential recorded movement. As such, the time span used to define the entries into and the exits from a state may penalize scholars in areas with a typically lengthy publication process. Therefore, using the mid-point between the year of change in the state of affiliation, as mentioned in section \ref{ss:modestate}, incorporates the assumption that movements to new affiliations are costly, and that the publication process is lengthy. We use a one-year net migration rate, $NMR_{i,t}$, of movements of scholars from state $i$ in year $t$:

\begin{equation}
\label{eqn:mig_rate}
NMR_{i,t}=\frac{I_{i,t}-E_{i,t}}{N_{i,t}} \times 1000
\end{equation}

Overall, the net migration rates are calculated as the difference between the entering ($I_{i,t}$) and the exiting ($E_{i,t}$) populations in state $i$ between time $t$ and $t-1$, as a share of the total population of scholars ($N_{i,t}$) in the state at the given time $t$. Scopus data allow for the measurement of the numerators and denominators for the net migration rates of scholars. In the absence of Scopus data, assuming that the scholars’ movements were similar to those of the general population would be a serious limitation.

We included several additional modifications for the components of this rate. As we mentioned previously, the numerator $t$ is the year corresponding to the mid-point between the change in states of a scholar’s movement history. We removed observations for which the total number of movements (entries and exits) fall within the lowest $15\%$ of observations, which removed cases that may not be representative relative to the scholar population size. The denominator is the aggregate count of scholars used for Figure \ref{fig:scholar_pc}, which uses a two-year backward and forward filled technique. As a result, the choice of filling locations limits the period for which migration rates can be computed to 1998-2016. Finally, from this point onward, all rates are expressed per 1,000 people.

\begin{figure}
 \centering
 \includegraphics[width=0.95\textwidth, scale=1.5]{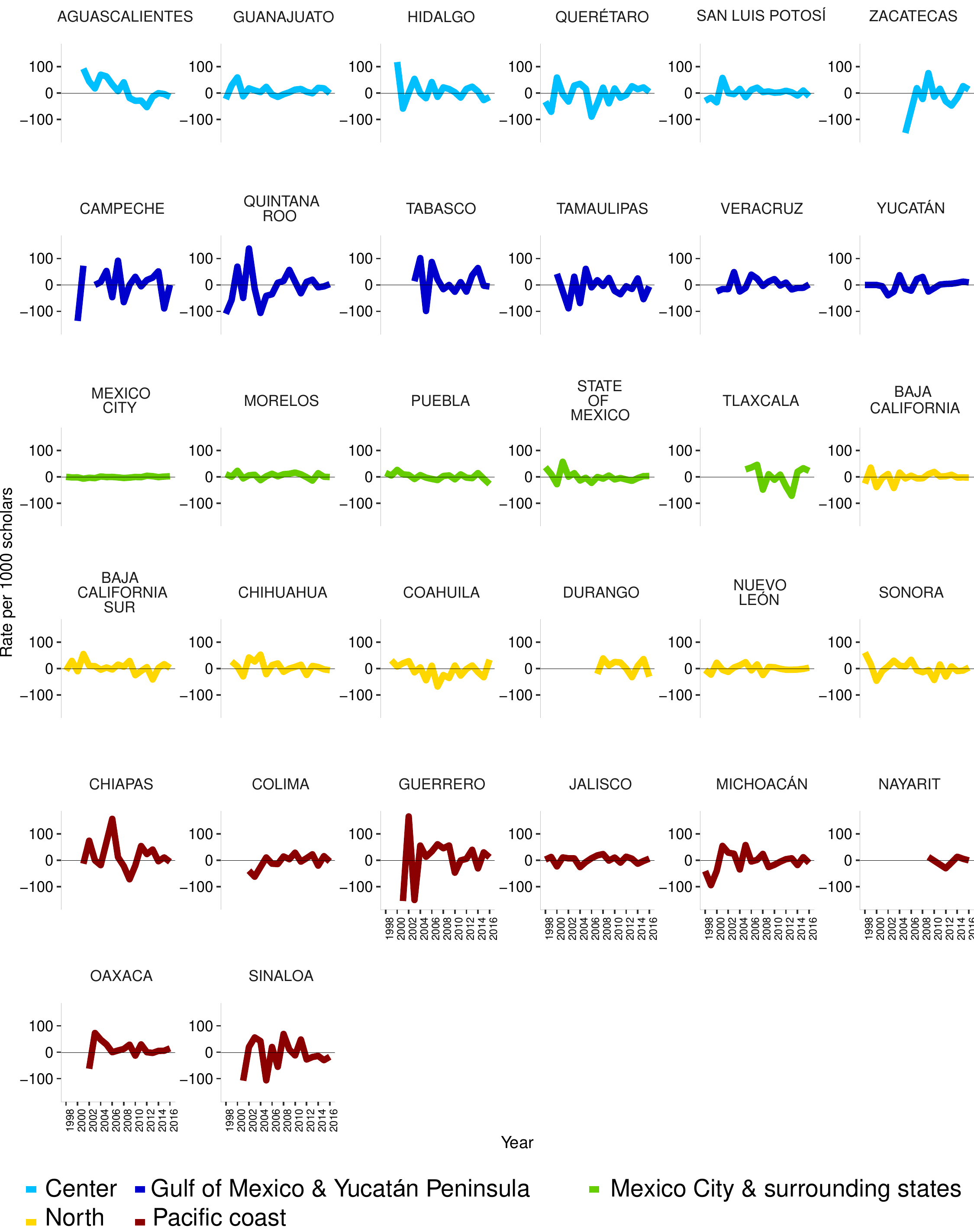} 
 \caption{Net migration rates for scholars by regions}
 \label{fig:net_migrate_states}
\end{figure}

Figure \ref{fig:net_migrate_states} illustrates the net migration rates of researchers in 32 states of Mexico. The net migration rates vary across states and within regions to the extent that we can start considering that there are underlying characteristics beyond geography that cause these disparities. Intuitively, a positive rate means that the entering researchers surpass those who left, while the opposite is the case for a negative rate. Rates close to zero correspond to situations in which there is no apparent loss or gain of scholars. That is, the number of scholars who leave is mostly offset by the number of scholars who enter. As academic institutions are bound by physical and budgetary constraints, there is a limit to the number of scholars who can be attracted and expelled. Unlike population growth rates, a consistently positive net migration rate of scholars may not be sustainable for some states.

We can observe three patterns experienced by the states: i) steady oscillations around zero, ii) downward or upward period-specific trends, or iii) large variation. When we consider the states of the Center, we see that in most of them, the rates fluctuate around zero (Guanajuato and San Luis Potos{\'i}); but that in others, such as Zacatecas and Hidalgo, the rates vary substantially. Consider the specific case of Aguascalientes. In recent decades, the state has become an industrial and technology hub that has attracted many universities and research centers. The initial downward but positive trend we observe in Aguascalientes may be the result of this industrial dynamism. However, with time, the growth in research opportunities may have decreased as the migration rate now approaches zero.

\FloatBarrier

When we turn to the states in the Gulf of Mexico and the Yucat{\'a}n Peninsula, we see that there are periods of substantial variation. Specifically, there are rates close to –100 scholars per 1,000 people. This means that there was a large-scale departure of scholars. These substantial variations coincide with the first few years of the Scopus data, which may be subject to a left-truncation such that our count of mobile and total scholars might be lower than the actual number. Therefore, the variation we observe may be an artefact of the lack of observed movements, so the results for the first few years of the analysis period should be interpreted with caution. 

On the opposite end of the spectrum, Yucat{\'a}n shows steady rates of around zero for the last years of the study period. We can focus on the states with the highest density of scholars as of 2016 from Figure \ref{fig:scholar_pc}: Mexico City, Morelos, Baja California Sur, Yucat{\'a}n, Baja California, Nuevo Le{\'o}n, and Quer{\'e}taro. Overall, their net migration rates fluctuate smoothly at around zero. Some of the lowest migration rates in absolute terms are found in Mexico City. It is possible that these states have already reached their capacity, but their institutions are still able to offer academic placements to scholars. Therefore, the zero net migration rates could imply that there is no brain drain in the states with the most scholars per capita.

Mexico City and its surrounding states, with the exception of Tlaxcala, together represent a notable case, as their rates are all close to zero. Beyond the researcher population, these states act as a large metropolitan area in which the economic and human flows are substantial. Unlike other regions, this area may be subject to overlapping characteristics that drive the similar migration rates. Conversely, the Pacific coast is an example of a region with large differences across states. Jalisco, Michoac{\'a}n, and Guerrero are neighboring states, but each has distinct migration rates. 

\begin{figure}[ht]
\centering
\includegraphics[width=0.95\textwidth]{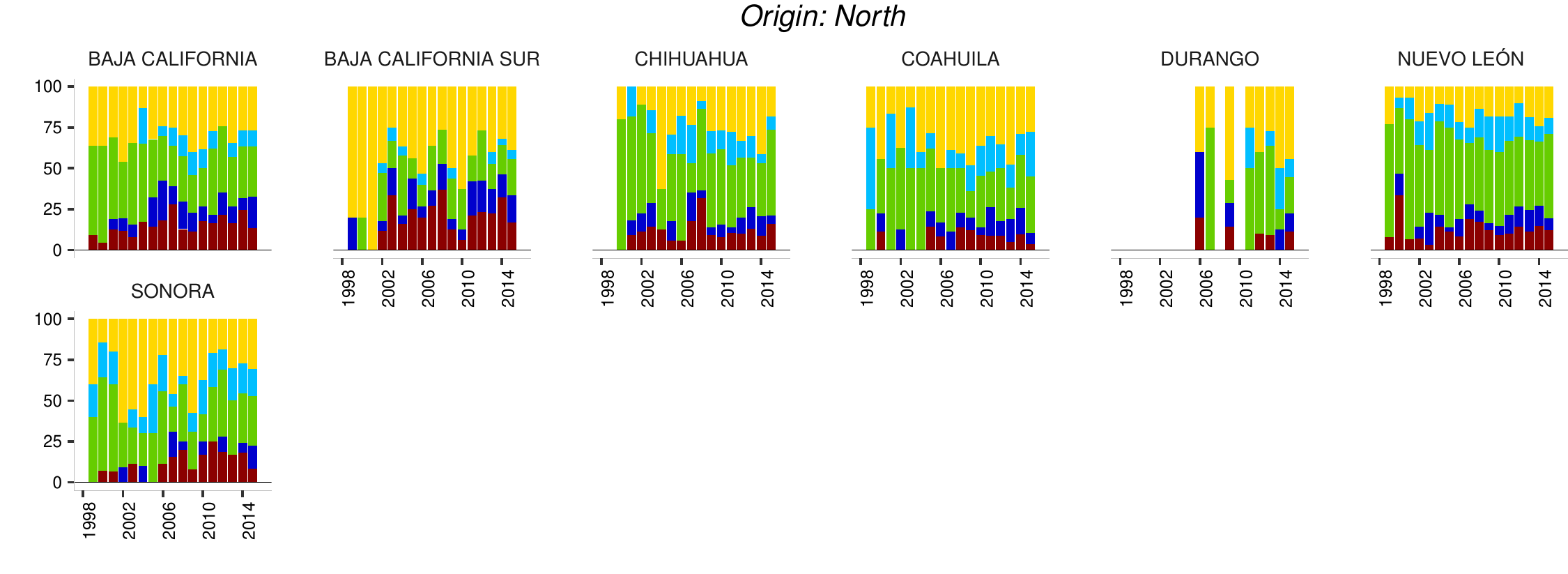} 
\includegraphics[width=0.95\textwidth]{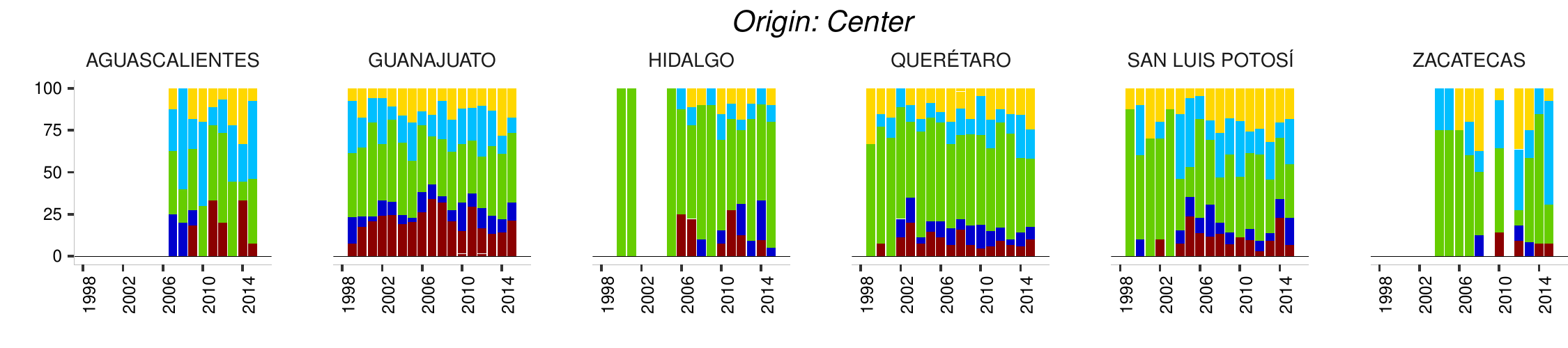} 
\includegraphics[width=0.95\textwidth]{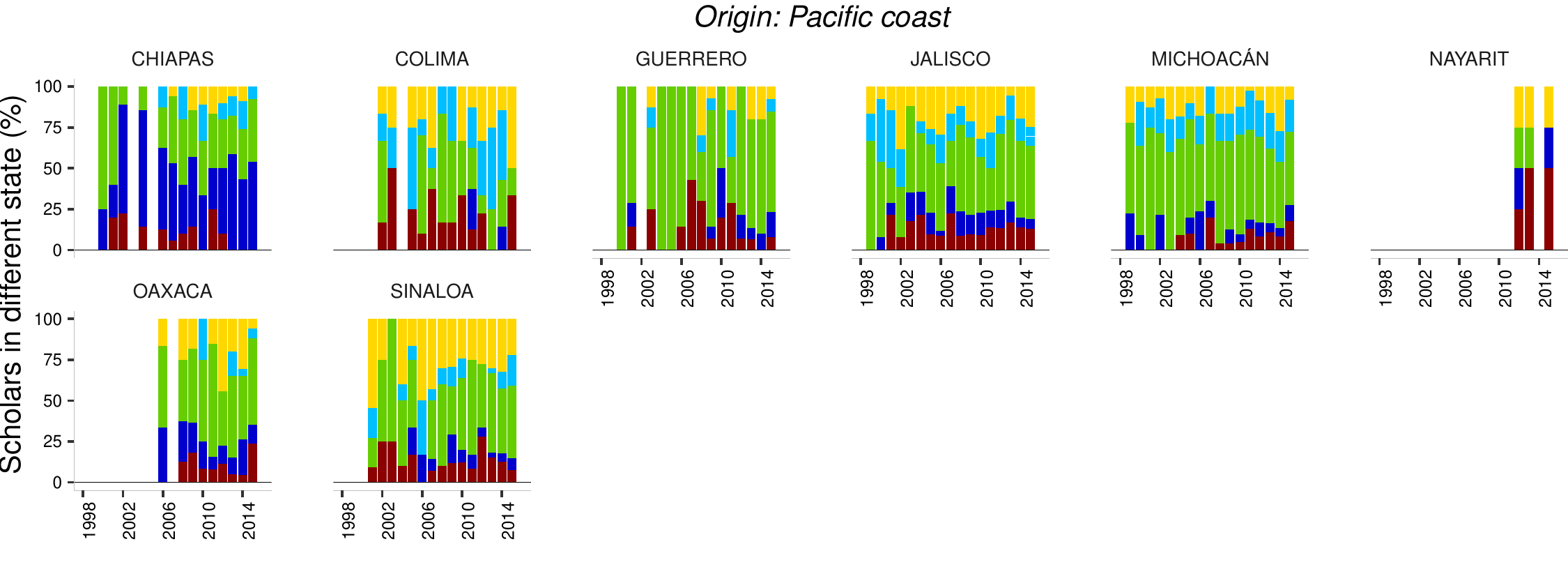} 
\includegraphics[width=0.95\textwidth]{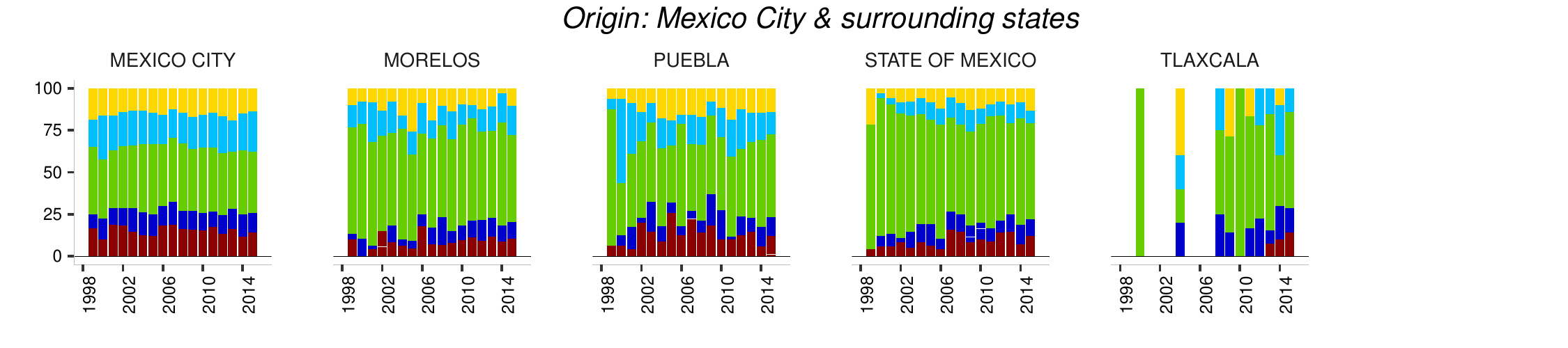} 
\includegraphics[width=0.95\textwidth]{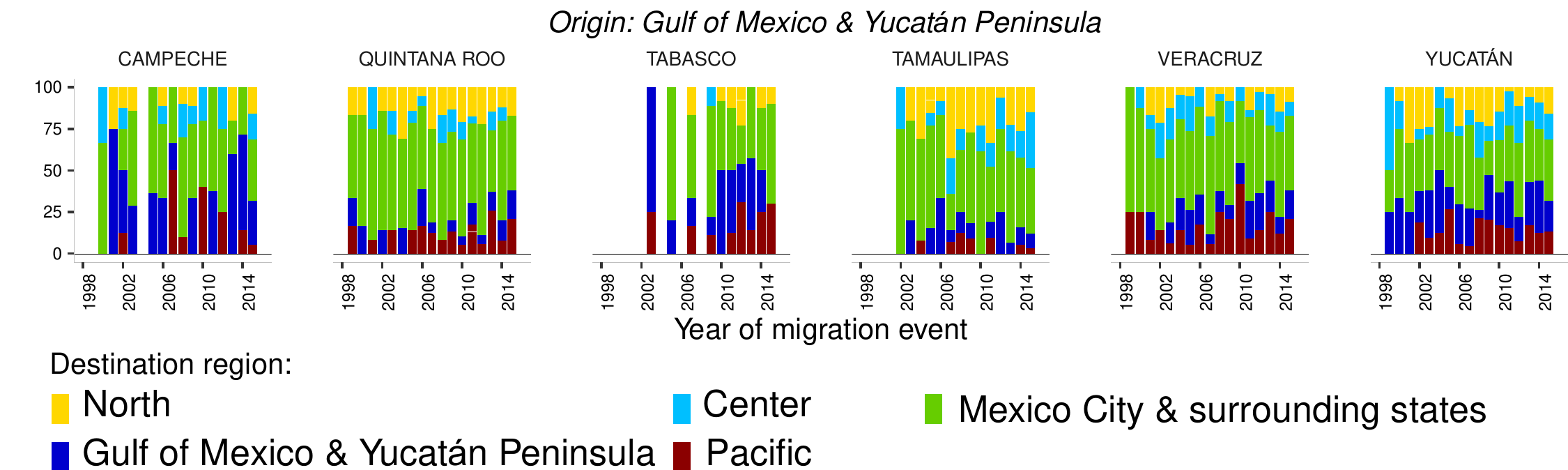}
\caption{Destination of mobile scholars in each year by their states of academic origin. Panels show for each state of origin the intensity of movements to other regions by year. Missing bars indicate years with a low number of migration events of scholars of a given origin.
}
\label{fig:origin_movements}
\end{figure}

Overall, our analysis of the net migration rates is an example of how bibliometric data can be used to detect details of the profiles of states that would not be visible when using alternative data, such as the census. Another perspective on the movements of scholars is to focus on changes relative to the state of the first publication. Figure \ref{fig:origin_movements} contains plots of the shares of mobile scholars who have moved to other states, by state of academic origin. Each row groups the states of academic origin within a given region. For instance, the first plot of Figure \ref{fig:origin_movements} shows that scholars who started off in Baja California mostly migrated to states within the Northern region, and to Mexico City and its surrounding states. There are cases in which the bars in the plot are missing; this occurs when there are few migration movements. Specifically, the lowest $15\%$ of the number of movements within a state per year are dropped from the chart. That is, the empty space is significant, as it indicates the states in which the migration of scholars is an extremely rare event. In states such as Durango, Aguascalientes, Zacatecas, Nayarit, Oaxaca, and Tlaxcala, the processed data can detect substantive movements for the last decade only.
\FloatBarrier

Figure \ref{fig:origin_movements} suggests that there is a degree of selectivity in the inter-state movements in a given state of origin. Scholars whose academic origins are in the Northern states tend to move within the region, with the exception of Nuevo Le{\'o}n. Although the Center states are surrounded by all of the remaining groups, it appears that scholars in the Center states are strongly attracted to Mexico City and the surrounding areas, as well as to the Pacific Coast (for the case of Guanajuato). Most of the mobile scholars who come from one of the states along the Pacific Coast or the Gulf of Mexico also prefer Mexico City and the surrounding states. However, since these regions have many neighboring states from other regions, there is some variation of preferences within regions. For instance, scholars from Chiapas seem to prefer going to the Gulf of Mexico region which is in line with the fact that it is the closest region other than Mexico City and the surrounding states. Similarly, Sinaloa is next to Durango, which belongs to the Northern states, and many of its originating scholars constitute the yellow bars representing the Northern states.

The least diverse outcomes are observed for the mobile scholars who began their career in Mexico City and surrounding states, as most move within the region. Once more, a potential explanation for this finding is that economic and political activities are centralized in Mexico City, and these developments have caused spillover connections to the neighboring states. Unlike for all other states, the shares of scholars moving to each region who began their career in Mexico City are similar. That is, even if scholars originated in Mexico City, they tend to also move to other regions, which implies that the destinations of those with their origins in Mexico City are not concentrated within one region. This origin-destination analysis suggests that regardless of the scholars’ state of origin, Mexico City and its surrounding states are preferred destinations.

The flexibility of the bibliometric data allows us to calculate additional migration rates to infer changes in the intensity and the redistribution effects of internal migration. First, the Crude Migration Intensity (CMI) measures the share of migration events relative to the population size \cite{bell2002cross}. Then, the Migration Effectiveness Index (MEI) and the Aggregate Net Migration Rate (ANMR) \cite{bell2002cross}, measure the effect of migration on the redistribution of a population within the country. Formally,

\begin{align}
CMI_{t} &=100 \times \frac{M_{t}}{\sum_{i}N_{i,t}} \label{eqn:cmi_rate}
\end{align}

\begin{equation}
\label{eqn:mei_rate}
MEI_{t} = 100 \times \frac{\sum_{i}|I_{i,t} - E_{i,t}|}{\sum_{i}(I_{i,t} + E_{i,t})}
\end{equation}

\begin{align}
ANMR_{t} &=100 \times \frac{0.5\sum_{i}|I_{i,t} - E_{i,t}|}{\sum_{i}N_{i,t}} \\ 
&= \frac{CMI_{t} \times MEI_{t}}{100} \label{eqn:anmr_rate}
\end{align}

\noindent where $i$ denotes a region and $t$ is the year of the movement\footnote{We extend the time-invariant measure in \cite{bell2002cross} into a time series.}. The numerator of the MEI contains the total sum of differences between the number of scholars entering ($I_{i,t}$) and exiting ($E_{i,t}$) a given region $i$. The MEI measures the net migration balance in an area as a share of the total number of individuals who moved either from, or away from, the zone. When the MEI is large, for every movement, there is a large contribution of the net migration rate, which means that migration is efficient in redistributing the population of scholars. While the ANMR has the same numerator as the MEI, it is weighted by the total population of scholars ($N_{i,t}$), and it can be expressed in terms of the CMI and the MEI. By changing the denominator to the population of scholars, the ANMR becomes an indicator of the general effect of population redistribution of moving scholars on the population, as shown in its second specification. Finally, due to our data format, the CMI is computed from the ANMR and CMI by rearranging equation \ref{eqn:anmr_rate}. Otherwise it would be obtained by the number of migration events ($M_t$), including those that enter and leave, over the total population at risk.

\begin{figure}[ht]
\centering
\subfloat[Crude Migration Intensity]{\includegraphics[width=0.45\textwidth]{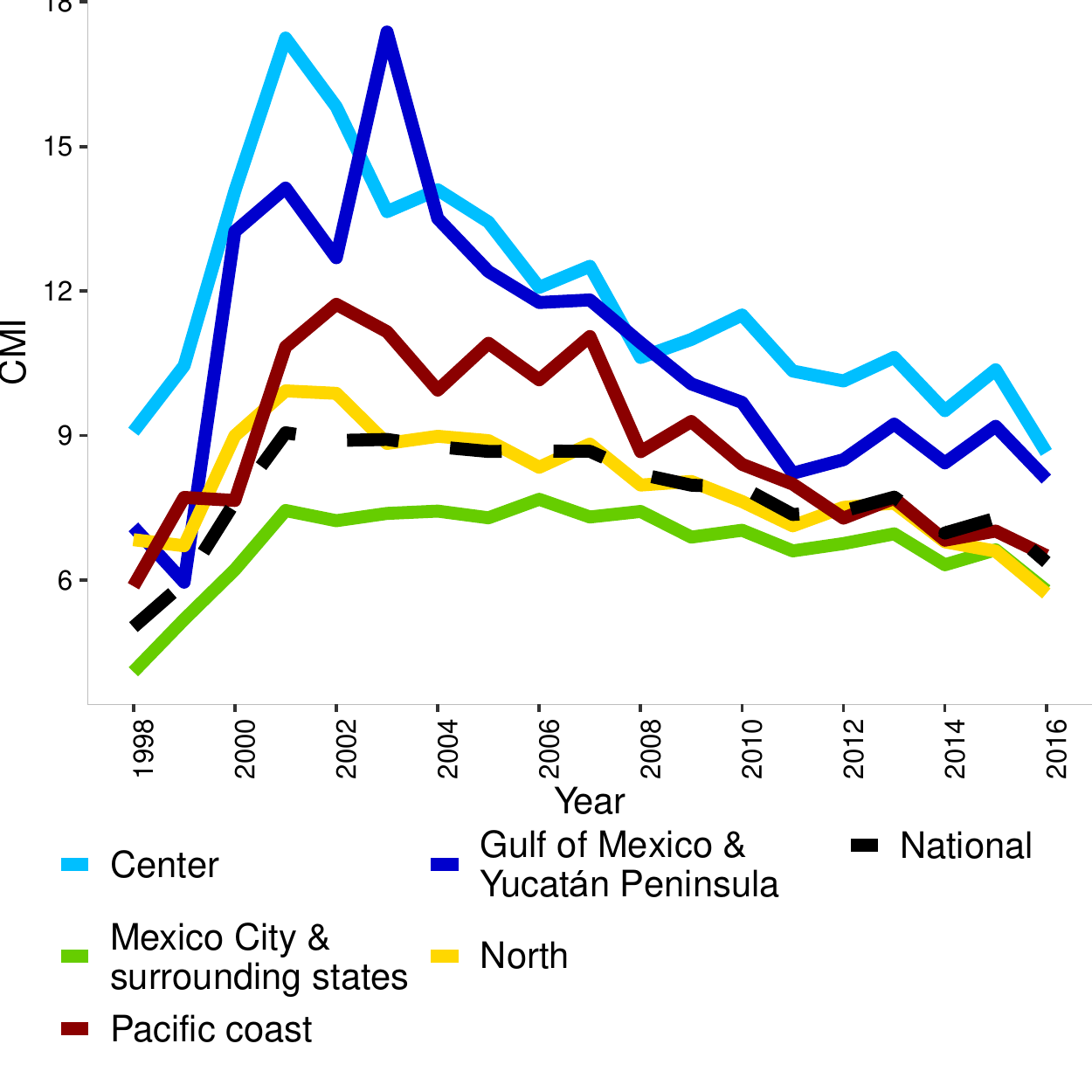} \label{fig:cmi} }

\subfloat[Migration Effectiveness Index]{\includegraphics[width=0.45\textwidth]{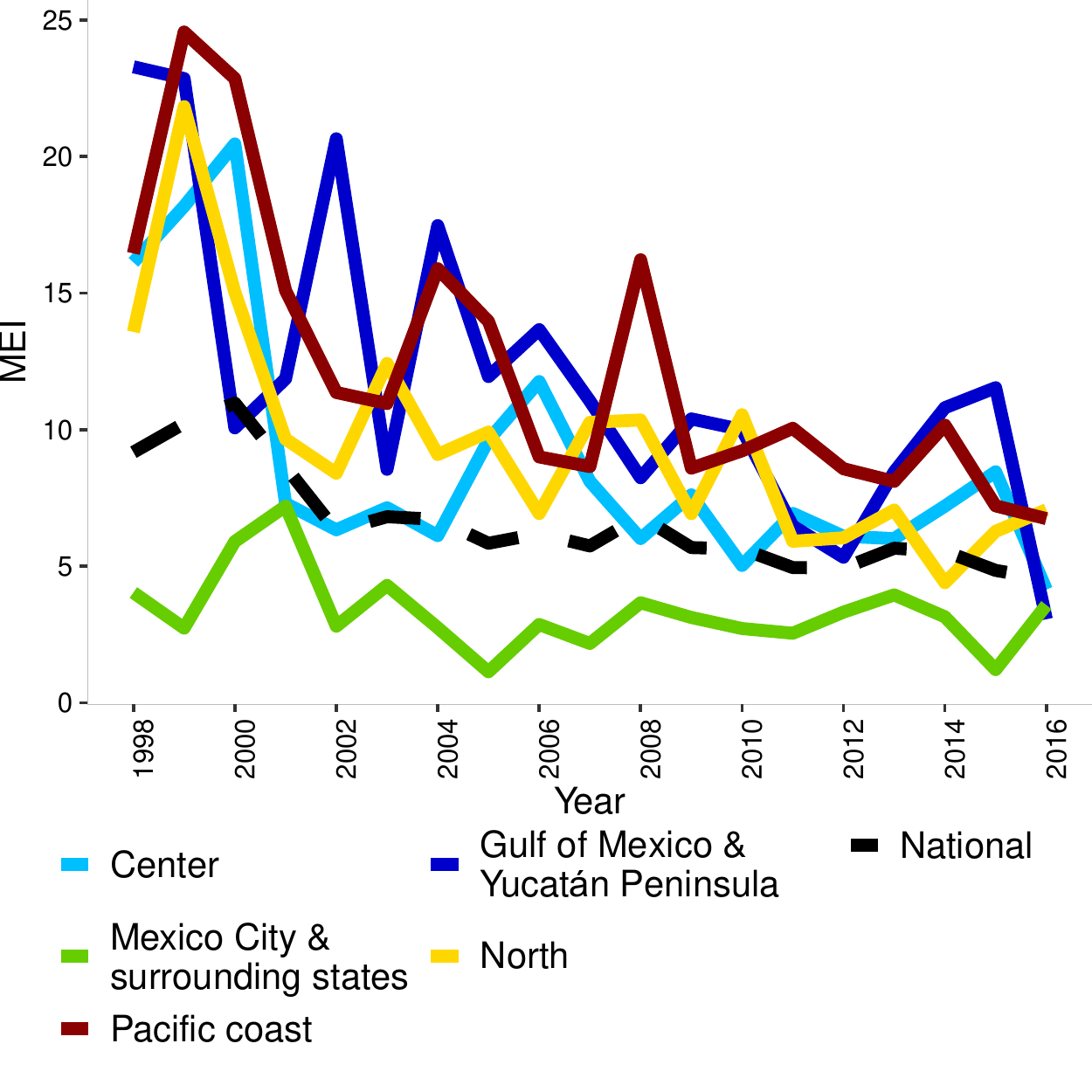}\label{fig:mei} }
\subfloat[Aggregate Net Migration Rate]{\includegraphics[width=0.45\textwidth]{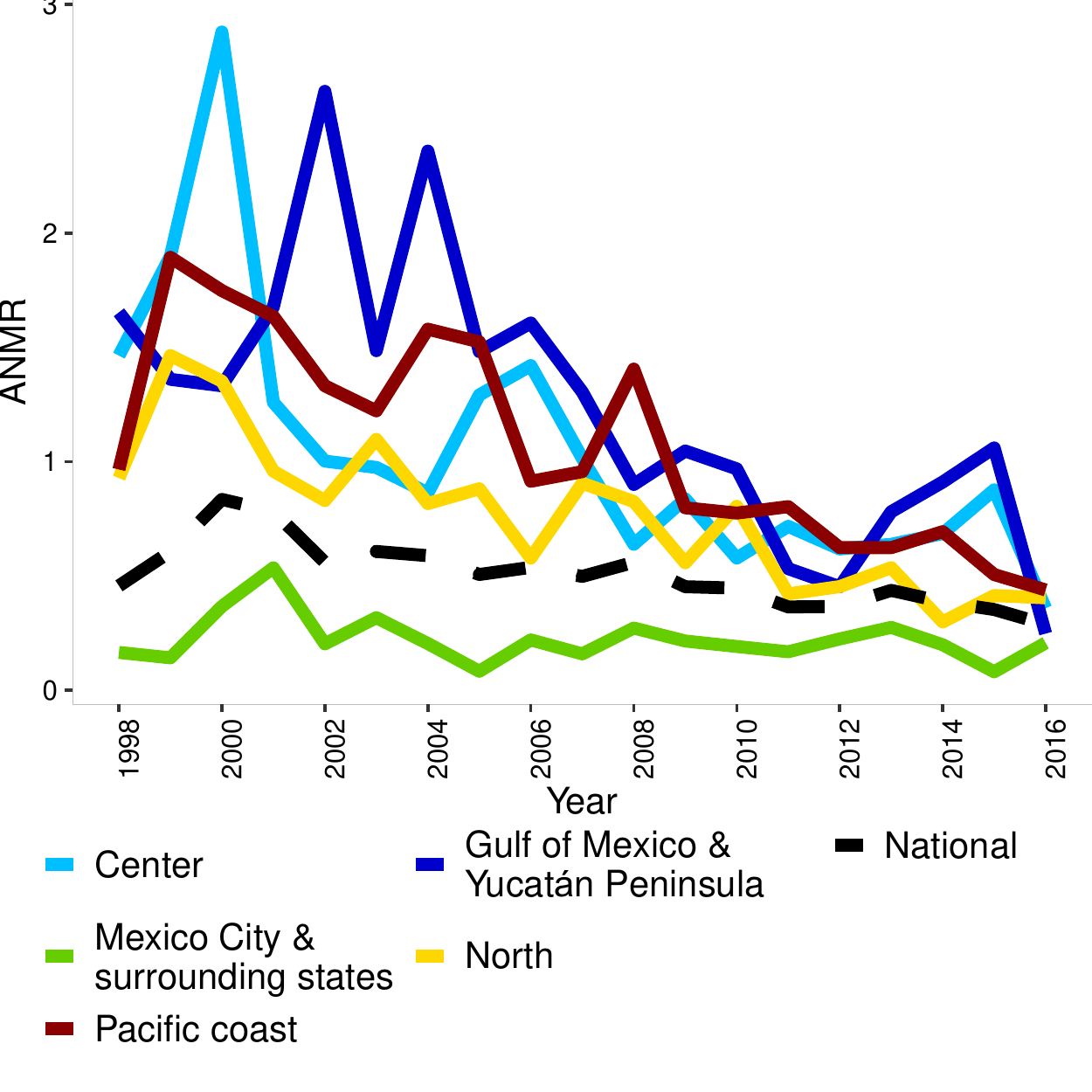}\label{fig:anmr} }
\caption{Migration intensity indices, over time. Dashed line represent country indices.}
\label{fig:migration_intensity}
\end{figure}

Figure \ref{fig:migration_intensity} shows values of the three indicators for mobile scholars for the 1998-2016 period. The results of the CMI measure displayed in Figure \ref{fig:cmi} shows that the country intensity is between five and nine, and has a decreasing trend. However, when we look at the intensity within regions, we see that the Center and Gulf regions experience the largest intensities. In other words, these regions seem to be more dynamic than the rest of Mexico as they experience more migration events after standardizing by their population size. This finding is an alternative reading of the results displayed in Figure \ref{fig:origin_movements}, which focused only on the shares relative to the origin state, rather than movements within the region regardless of the origin. \footnote{For example, the North CMI is composed of the yellow bars of states that belong in the Northern region, and of the yellow bars from researchers who did not start off in this region but eventually moved within it.} Mexico City and the surrounding states have the lowest regional CMI, which suggests that controlling for scholar population size reduces the regional importance of mobile scholars. While Figure \ref{fig:origin_movements} shows that many scholars prefer Mexico City and the surrounding states, their contribution seems to diffuse relative to all scholars within the region. 

In general, the country MEI in Figure \ref{fig:mei} shows a downward trend, which suggests that the movements of scholars are less efficient over time. When we examine the within-region migration efficiency of scholars, we observe that the MEI measures are highly volatile, except those for Mexico City and its surrounding states. In comparison to the rest of the states, the low MEI from the Mexico City area suggests that there is a less efficient movement of scholars within this region.

As expected, the results for the ANMR measurements shown in Figure \ref{fig:anmr} are similar to those in the MEI. However, the ANMR values suggest that the turnover of scholars as a share of the population is much lower and stable for Mexico City and the surrounding states. Overall, these results jointly suggest that the net migration level to the Mexico City area is lower than the total volume of mobile scholars in that region, and, therefore, that the level of population redistribution is low.

The results presented throughout the section highlight the versatility of bibliometric data for demographic analyses of scholars. By re-purposing the data, we can create profiles for states based on their scholar population and their movements. Even when there is heterogeneity within the regions, states with higher scholar density tend to have migration rates around zero, which suggests that there is a brain \textit{balance}, rather than gain or drain. However, even when we use intensity measures, we cannot analyze between-region patterns from a demographic standpoint. The following section contains a network analysis that provides a complementary perspective to help us better understand the mobility of scholars.
\FloatBarrier

\section{Internal migration as a network}
\label{s:network}

The data from scholars who have moved between different states can be used to create a migration network by representing each state as a node, and creating a directed edge $(i,j)$ for each mobility event from state $i$ to state $j$. Creating such a migration network enables us to analyze the system of scholarly internal migration in Mexico as a whole. We have made the data publicly available in a \textit{FigShare} data repository \cite{miranda_gonz_data_2020}.

\subsection{Networks and their temporal dynamics}

\begin{figure}[ht]
\centering
\subfloat[1996-2018, edges representing movements of 5 or fewer people are not shown
]{
\includegraphics[width=0.45\textwidth]{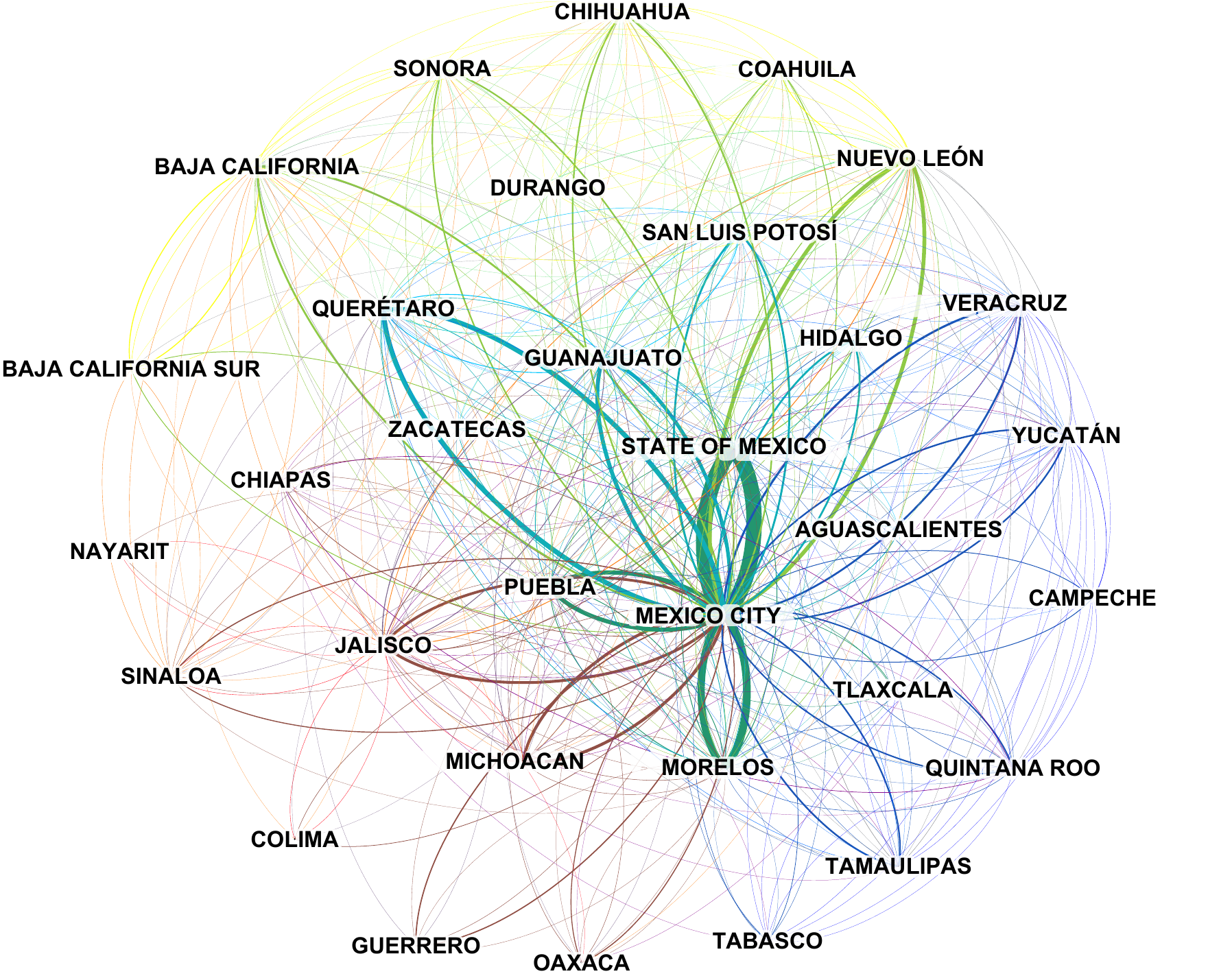} 
}
\subfloat[1997-1998]{
\includegraphics[width=0.45\textwidth]{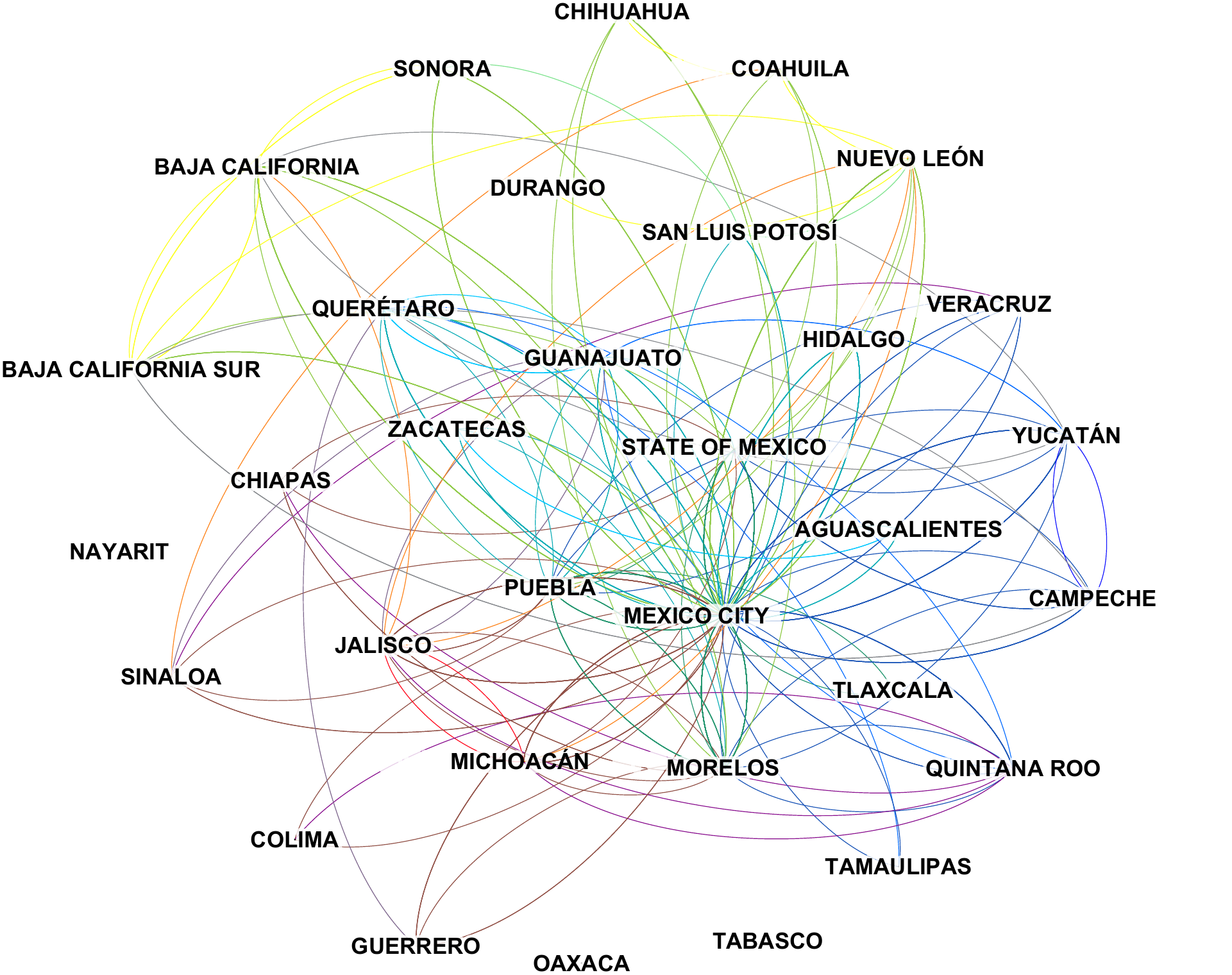} 
}
\hfill
\subfloat[2007-2008]{
\includegraphics[width=0.45\textwidth]{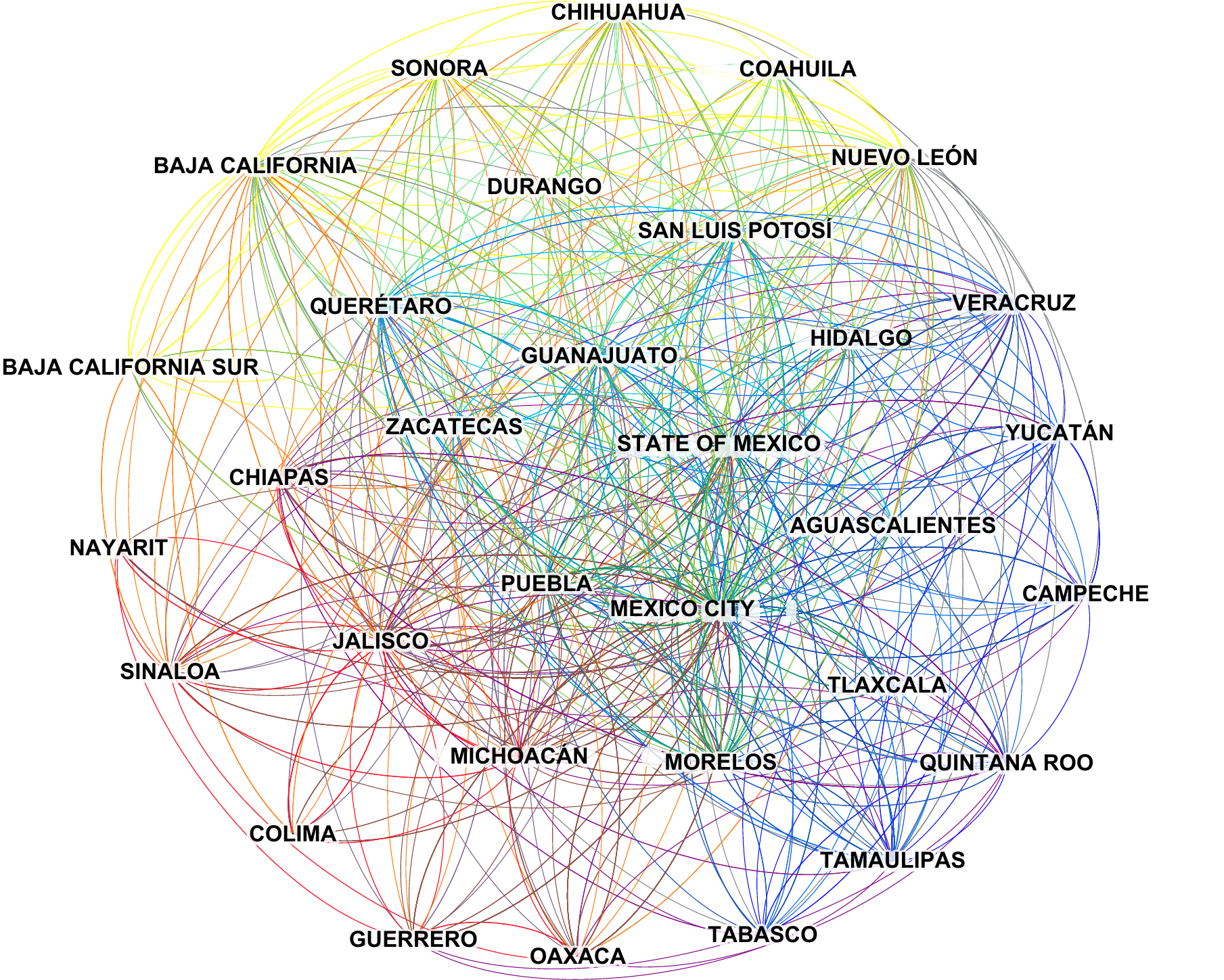} 
}
\subfloat[2017-2018]{
\includegraphics[width=0.45\textwidth]{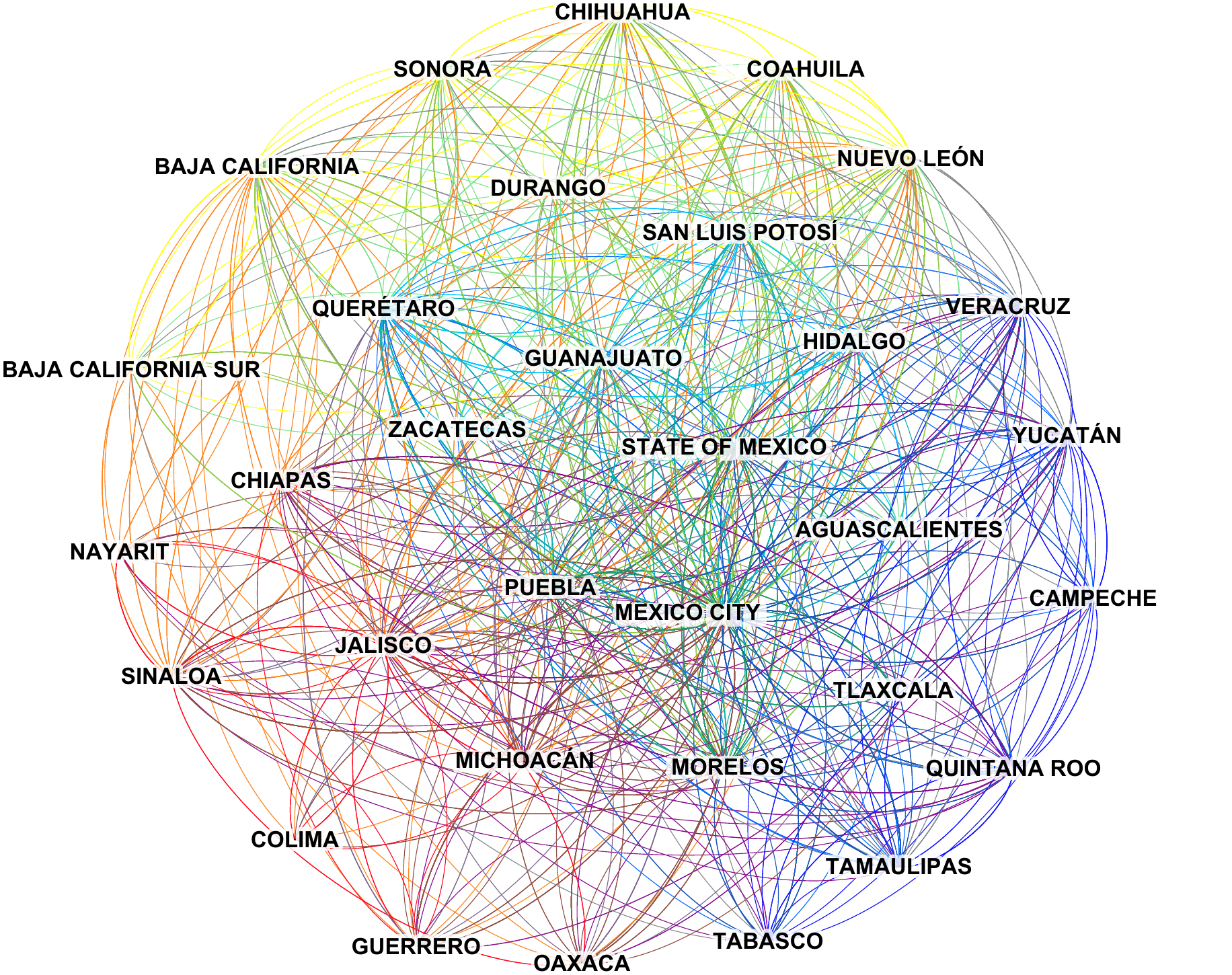} 
}
\caption{Network of internal migration among researchers in Mexico in 1996-2018 in which intensity of movements is shown by the thickness of the edges (a), three cross-sectional networks based on selected one-year periods (b-d). Directions of edges are clock-wise and their colors are the mix of respective origins and destinations (see the figure on screen for high resolution).}
\label{fig:network}
\end{figure}

Figure \ref{fig:network} (a) shows the direction and the magnitude of migratory movements of scholars in Mexico between 1996 and 2018. The five states that send the most scholars include Mexico City, the State of Mexico, Morelos, Nuevo Le{\'o}n, and Guanajuato, respectively (these are also the states that receive the most scholars, and in the same order, except that Guanajuato outranks Nuevo Le{\'o}n). These five states represent the capital city and its surrounding states, as well as the states that contribute the most to national GDP. As it can be seen in panel (a) of Figure \ref{fig:network}, the two flows between every pair of states are mostly reciprocal, with similar intensities (which means that the adjacency matrix of the directed network is very close to being a symmetric matrix). Aggregating the flows for each node, we see a similar pattern in node degrees. The in- and the out-degree of the same nodes are strongly correlated (the Pearson correlation coefficient is 0.999).

Mexico City has the highest incoming and outgoing flows of mobile scholars; i.e., its flows are almost three times as large as the flows in the State of Mexico, which is the second degree central node of the network. This is likely due to Mexico City’s political and economic importance, and because it is home to many large national universities and research institutes. There is substantial heterogeneity in the degrees of the nodes. The states with the lowest in- and out-degrees are Nayarit, Tlaxcala, Colima, Durango, and Zacatecas (in increasing order of degrees); their in- and out-flows are less than 0.02 of those of Mexico City.

Subpanels (b-d) of Figure \ref{fig:network} highlight the period movements of scholars between states. Overall, the migration network of researchers has not only become more dense (from a density of 0.08 in 1996-1997 to 0.5 in 2017-2018), but its edges have also become more diverse over the past two decades. For instance, in more recent years, the states along the Pacific coast (red) have had a greater exchange (purple edges) with states along the Gulf of Mexico and the Yucat{\'a}n Peninsula (blue).

\subsection{Assortativity of networks over time}
The degree assortativity of a network captures the correlations between the degrees of adjacent nodes \cite{newman2003mixing}. In many social networks, there is a tendency between nodes of a similar degree to connect (assortative mixing by degree); whereas in many technological and biological networks, high degree nodes tend to connect with low degree nodes (dissortative mixing by degree). Moreover, in some other networks, there is no correlation between degrees of adjacent notes (random mixing pattern) \cite{newman2002}. In directed networks, the correlation can be measured in four different ways by using either the in- or the out-degree for the source and target nodes (in-in, in-out, out-in, and out-out). We measure the out-in assortativity of directed graph $G$ using Eq.\ \eqref{eq1} \cite{newman2003mixing}.

\begin{equation}
 \label{eq1}
 r(G)=\frac{\sum_{j,k}(e_{j,k}-q_j^\text{in}q_k^\text{out})}{\sigma_\text{in}\sigma_\text{out}} 
\end{equation}

In Eq.\ \eqref{eq1}, $r(G)$ is the out-in degree assortativity coefficient for directed graph $G$. $e_{j,k}$ is the probability that a randomly chosen directed edge leads into a vertex of in-degree $j$ and out of a vertex of out-degree $k$. The term $q_j^\text{in}$ ($q_k^\text{out}$) represents the excess in-degree (out-degree) distribution \cite{newman2003mixing} of directed graph $G$ %i.e., the probability that a randomly chosen vertex has in-degree (out-degree) equal to $k+1$, 
and $\sigma_\text{in}$ ($\sigma_\text{out}$) is the standard deviation of the distribution $q_j^\text{in}$ ($q_k^\text{out}$). 

\begin{figure}[ht]
 \centering
 \includegraphics[width=0.6\textwidth]{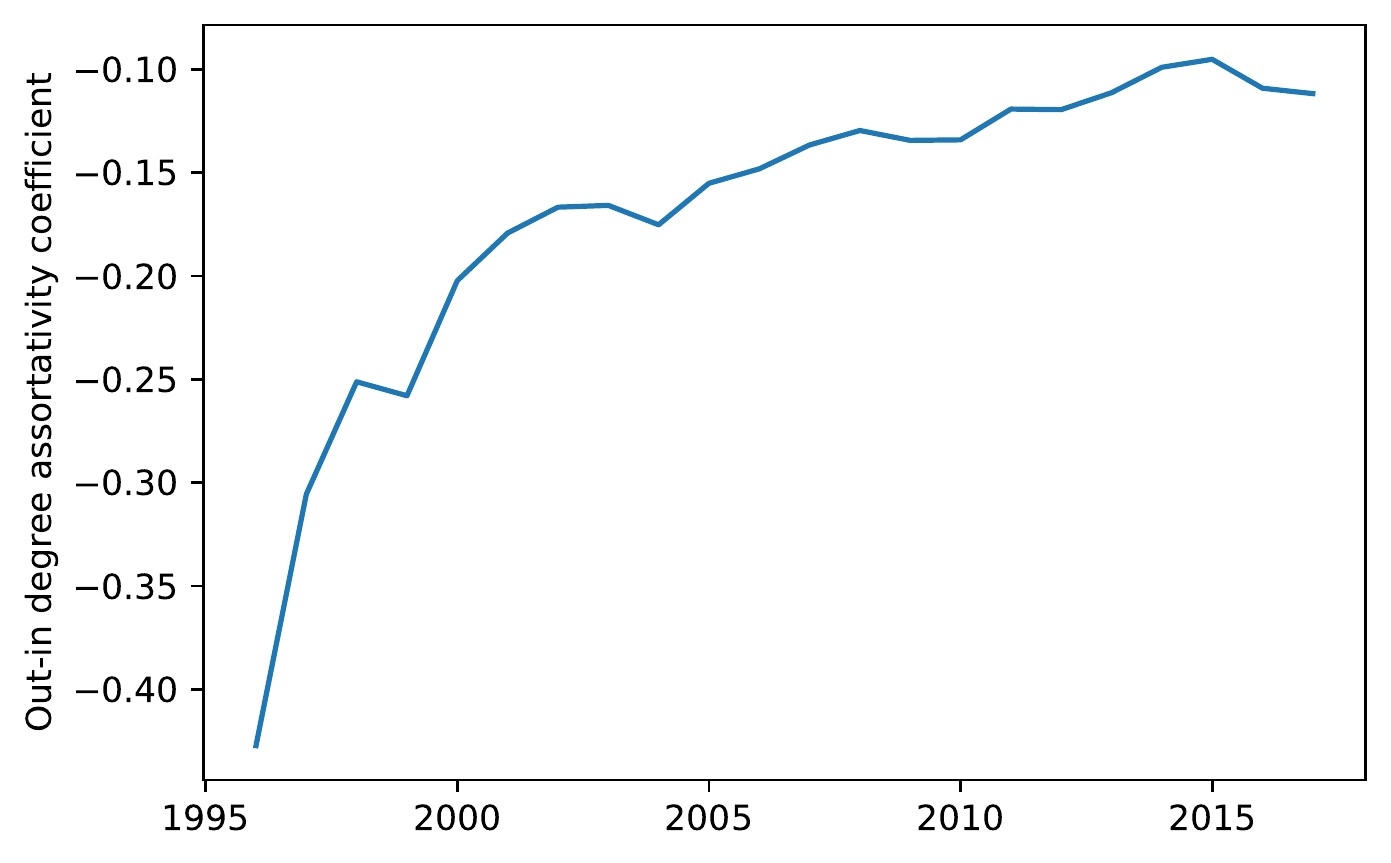}
 \caption{The out-in degree assortativity of the cross-sectional one-year networks showing an overall increase over time}
 \label{fig:assortativity}
\end{figure}

Figure \ref{fig:assortativity} illustrates $r(G)$ for the cross-sectional one-year migration networks. As it can be seen in Figure \ref{fig:assortativity}, the degree assortativity coefficient of the networks has generally increased over time during the 1996-2018 period. This increase in assortativity can be explained as the gradual transformation of the network from a dissortative mixing pattern ($r(G)=-0.45$) to a random (less dissortative) mixing pattern ($r(G)=-0.1$).

In the first few years of the period under study, movements were mostly from low out-degree nodes (states with small outgoing flow) to high in-degree nodes (states with large incoming flow), and from high out-degree nodes to low in-degree nodes. Given the strong correlation between the in- and the out-degrees of the same nodes, all four types of degree assortativity measures show similar increases, albeit with minor differences (in-in, in-out, and out-out plots are not shown to avoid redundancy). Therefore, we can say that the majority of flows in the first few years were from states with low flows to states with high flows, and vice versa.

However, over the last few years, the degrees of adjacent nodes are hardly correlated. The mixing patterns of the networks in more recent years display less dissortativity, and instead indicate that researchers are circulating between the states, irrespective of their degrees. This observation suggests that the mixing pattern of the network is affected by a relative increase in mobility between states with similar flow intensities (low-flow to low-flow and high-flow to high-flow) in the past two decades.

\subsection{Community structure induced by the dynamics of flows}

To detect communities in networks, different approaches (and algorithms) can be used. According to Fortunato and Hric \cite{FORTUNATO2016}, these approaches belong to five categories: optimization, statistical inference, dynamics, consensus clustering, and spectral methods. As the edges in our migration networks represent flows between states, we expect that detecting and analyzing communities based on the dynamics of the flows may reveal a more meaningful structure. \textit{InfoMap} is a popular algorithm for detecting communities from the dynamics category \cite{Rosvall2008}, which relies on random walk dynamics in the network, and on the intuition that a hypothetical random walker stays in the dense regions of the network for a longer period of time. For running InfoMap, we use the \textit{MapEquation} software package \cite{Rosvall2010,edler2013mapequation} to detect how the flows between the states lead to the formation of network communities between them. The network communities could be thought of as different components of the Mexican scholarly migration system, which are revealed due to the heterogeneity of the flows.

\begin{sidewaysfigure}
 \centering
 \includegraphics[width=1\textwidth]{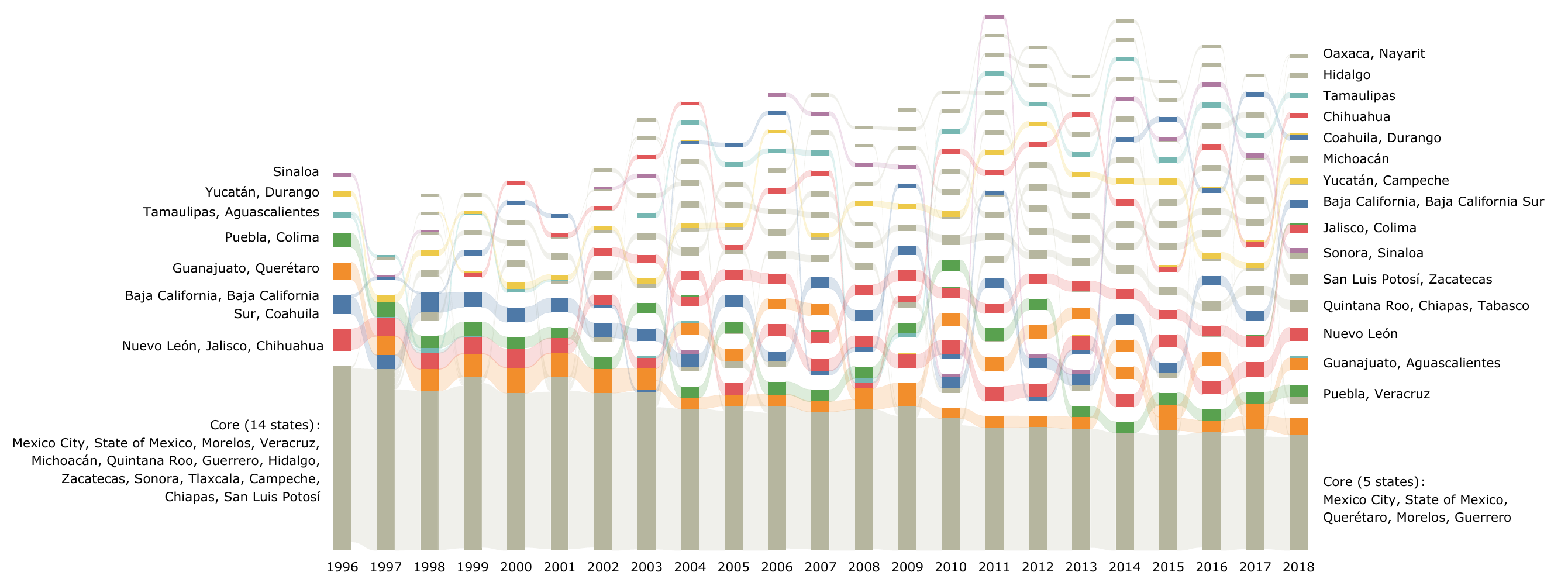}
 \caption{Communities of the states over time featuring a core-periphery structure}
 \label{fig:alluvial}
\end{sidewaysfigure}

For this purpose, we have created an alluvial map \cite{Rosvall2010} of the network flows and communities over time, which is illustrated in Figure \ref{fig:alluvial} ($99\%$ of the total flow is shown). The height of each community is proportional to its total flow. The structure of these communities clearly shows a dense core, with relatively few nodes accounting for the majority of the total flows, and an outlying, loosely connected periphery made up of a relatively large number of nodes accounting for a small portion of the flows. Such a network structure is referred to as a core-periphery structure \cite{borgatti2000models}.

As it can be seen in Figure \ref{fig:alluvial}, there were eight communities in the migration network in 1996. The core was made up of 14 states listed in the figure. The other communities in 1996 were smaller, and were made up of at most three states (as shown in Fig.\ \ref{fig:alluvial}). Most of the communities of 1996 have a dynamically changing flow over different years, and therefore break into smaller groups over time, which join back together in some years. For instance, the community shown in dark blue is made up of Baja California, Baja California Sur, and Coahuila in 1996. This community breaks into different groups the next year, and then reforms to its previous state one year later. By 2018, Baja California and Baja California Sur are together in a two-node community.

As the dynamics of flows in some other communities are more volatile, their nodes may break off at some point, and never again form a community. For instance, Nuevo Le{\'o}n, Jalisco, and Chihuahua formed a three-node community in 1996, which is shown in dark red. The three nodes then diverge into different groups. Eventually, in 2018, each of them is a part of a different community, possibly with other states. In 2018, the number of communities increases to 16, and the core community becomes smaller, both in terms of the relative share of the annual flow (represented by height) and the number of nodes, which is reduced to five (states are listed in the figure).

\section{Discussion and Summary}
\label{s:discuss}

This project was undertaken to analyze the internal migration of researchers within a country, and to enhance and evaluate the potential suitability of bibliometric data for such an analysis. After investigating what is achievable within the limits of classic sources of migration data, such as surveys and censuses, we outlined a detailed methodological framework for re-purposing bibliometric data for studying internal migration. 

Our framework involves an author name disambiguation method that enhances the quality of author identification offered by the provider of bibliometric data, and ascertains its accuracy in identifying the publications of the same author by treating the outliers. An initial step of our proposed method involves a parsimonious rule-based algorithm that infers states from affiliation addresses. The output of this step, combined with manually tagged data, is then used to train a neural network for predicting the states for the rest of the data. The produced state variable is used, as the sub-national level of data aggregation on which the locations of researchers are inferred over a number of years. The final step of our methodological framework deals with detecting migration events from changes in modal states of affiliations. While we illustrated this methodological framework for the case of Mexico, this approach is applicable to a broader context.

In an effort to bridge the gap between demographic methods and the literature on the migration of scholars, we estimated key internal migration measures over time, such as the net migration rate, the migration effectiveness index, and the aggregate net migration rate. To the best of our knowledge, these types of estimates have not previously been produced using bibliometric data. Substantively, we provided a detailed picture of scholarly migration within Mexico, we showed patterns of movements by origin and destination over time, and we offered comparisons of the states and regions with respect to scholarly migration.

We also modeled the movements of researchers as a network, which enabled us to analyze them holistically, and to evaluate their progression through time. Using the degree of the nodes, we identified the symmetry of pairwise flows of opposite direction, and the heterogeneity that exists at the level of flows between the states. Using degree assortativity, we explained the temporal changes in the correlations of the in- and the outflows between the states, and how the movements between states with similar total flows have increased relatively over time. We documented a core-periphery structure in the network, and explored its dynamics through detecting network communities induced by the changes in the flows between states. The analysis revealed a system whose core community has decreased in both size and total flow over time, while smaller peripheral nodes gained a larger share of the network flow and formed their own communities.

The changes observed in patterns of scholarly migration between states can be viewed from the perspective of a theoretical model like Zelinsky’s migration transition model \cite{Zelinsky1971} introduced in Section \ref{s:intro}. Although Zelinsky’s model is based on a general population of migrants, it may have a bearing on mobile sub-populations such as scholars, as we argue that their mobility could change with respect to the developmental stages of the society. The initial patterns of migration between rural and urbanized states for the first few years of the period under study (Figure \ref{fig:network} (a)) was followed by an increase in assortativity, under which flows between nodes of similar degrees (including high-degree nodes) grew in relative terms (Figure \ref{fig:assortativity}). This suggests that mobility among scholars in Mexico might have been experiencing a phase similar to Zelinsky’s \textit{late transitional society} \cite{Zelinsky1971}. The assertion of the theory is that at this stage, there is a relative increase in migration between urban centers. This, in turn, results in circular migration within a single metropolitan region of the network. With our current analysis, we cannot test whether broad mobility theories, like the migration transition model, also apply to scholars. Similarly, providing conclusive evidence on how the trajectory of scholarly migration is similar to or different from the experience of the general population is beyond the scope of this article. However, we believe that our analysis, which is mostly empirical in nature, lays the foundations for testing and developing theories that can rely on the analytical framework developed by migration scholars, and the richness of appropriately processed bibliometric data.

By offering novel methodological contributions, our work demonstrates that large-scale longitudinal bibliometric data could be harnessed to obtain valuable insights into the internal migration patterns of scholars when coupled with an algorithmic method for producing a sub-national level for data aggregation. Among other contributions, we documented (i) that mobile scholars, on average, publish more and have higher citation counts than non-mobile scholars; (ii) that there is substantial heterogeneity in net migration rates across states, but that Mexico City and the surrounding states maintain a position of centrality as key areas that attract scholars from all regions of Mexico, and send scholars to all states; (iii) that, overall, the crude migration intensity and the migration effectiveness index for scholars have declined over time, which could indicate that the redistribution of scholars in Mexico has decreased in the past two decades.

More broadly, we leveraged demographic techniques to provide extensive results on scholarly migration in Mexico, a topic that has been understudied despite the considerable size and impact of the country’s research. The methodological framework proposed in this study aims to facilitate organizing data and information about the migration of scholars that can be used to evaluate the likelihood of alternative hypotheses, and to build the foundations of a theory of scholarly migration.

While standard demographic measures, such as net migration rates, are essential for quantifying migratory movements, a more comprehensive picture of scholarly migration is obtained by additionally drawing from network approaches. Demographic and network approaches complement each other in providing a more comprehensive view on the dynamics of scholarly migration that is consistent with the transitional nature of migration systems. For example, while we observed a general decreasing trend in the crude migration intensity over the past two decades, we also found that the migration network has become more dense and more diverse, and has been characterized by increased exchanges between the states along the Gulf and the Pacific Coast. The combination of methods and data that we present opens up new opportunities for developing a theoretical framework for understanding scholarly migration within country boundaries.

\section*{Declarations}

\subsection*{Availability of data and material}
The bibliometric data used in this study is proprietary and cannot be released. Scopus data is owned and maintained by Elsevier. Our network dataset is made publicly available in a \textit{FigShare} data repository \cite{miranda_gonz_data_2020}.

\subsection*{Competing interests}
Authors declare no conflicts of interest.

\subsection*{Funding}
This study has received access to the bibliometric data through the project ``Kompetenzzentrum Bibliometrie" and the authors acknowledge their funder Bundesministerium für Bildung und Forschung (funding identification number 01PQ17001). The first author has received support in the form of a Doctoral fellowship from CONACYT and the University of California Institute for Mexico and the United States (UC MEXUS).

\subsection*{Authors' contributions}
AMG and SA did the majority of the work related to the research design, the data processing, the analysis, and the writing and editing of the manuscript. TT contributed to the development of the neural network and the author disambiguation algorithm. TT and EZ contributed to the writing and revision of the manuscript.

\subsection*{Acknowledgements}
The authors are grateful to the anonymous reviewers whose comments substantially improved this study. The authors highly appreciated the discussions with Fariba Karimi and Daniel Edler; the comments from Andre Grow, Denys Dukhovnov, Diego Alburez, Jorge Cimentada, Sophie Lohmann, and Daniela Negraia; and the data quality checks done by Jakob Voigt.

\bibliographystyle{bmc-mathphys}
\bibliography{references}

\end{document}